\documentclass[11pt,a4paper]{article}
\pdfoutput=1

\usepackage{jheppub}
\usepackage{rotating}
\usepackage{orcidlink}
\usepackage{wrapfig}
\usepackage{amsfonts}
\usepackage{amsmath}
\usepackage{slashed}
\usepackage{amssymb}
\usepackage{latexsym}
\usepackage{multirow}
\usepackage{hyperref}
\usepackage{enumitem}
\hypersetup{colorlinks=true,citecolor=red,linkcolor=magenta,urlcolor=blue}
\usepackage{relsize}
\usepackage{booktabs}
\usepackage{cleveref}
\usepackage{fancyhdr}
\usepackage{float}
\usepackage{graphicx}
\usepackage{microtype}
\usepackage{tabularx}
\usepackage{xcolor}
\usepackage[bottom]{footmisc}
\usepackage{soul}
\usepackage{tikz}
\usepackage{tikz-feynman}
\usepackage{babel}
\usepackage{subfigure}

\preprint{\texttt{DESY-26-079, 	RBI-ThPhys-2026-13}}

\title{Kinematic Riffs and Interference Effects in Triple Higgs Production in the N2HDM}

\abstract{
We investigate the complex kinematic structure involved in resonant production of three Higgs bosons at the Large Hadron Collider (LHC), within the rich scalar spectra of the Next-to-minimal Two Higgs Doublet Model (N2HDM), which features three CP-even neutral scalar degrees of freedom. Focussing on the resonant topologies, we analyse the invariant mass and transverse momentum distributions to disentangle the underlying production mechanisms. We demonstrate that interference effects and additional kinematically accessible decay channels can significantly alter kinematic observables, highlighting the limitations of simplified approximations and underscoring the importance of fully differential studies for probing extended Higgs sectors.
}
\author[a]{Wrishik Naskar\orcidlink{0000-0002-4357-8991},}
\author[b]{Tania Robens\orcidlink{0000-0002-9913-5225},}
\author[c]{Julia Anabell Ziegler\orcidlink{0009-0005-8817-0729}}
\affiliation[a]{Deutsches Elektronen-Synchrotron DESY, Notkestr.~85, 22607 Hamburg, Germany}
\affiliation[b]{Ruder Boskovic Institute, Bijenicka cesta 54, 10000 Zagreb, Croatia}
\affiliation[c]{II. Institut f\"ur Theoretische Physik, Universit\"at Hamburg, Luruper Chaussee 149, 22761 Hamburg, Germany}
\emailAdd{wrishik.naskar@desy.de}
\emailAdd{trobens@irb.hr}
\emailAdd{julia.ziegler@desy.de}
\begin{document}
\maketitle
\flushbottom

%%%%%%%%%%%%%%%%%%%%%%%%%%%%%%%%%%
\section{Introduction}
\label{sec:intro}
%%%%%%%%%%%%%%%%%%%%%%%%%%%%%%%%%%
As the Large Hadron Collider (LHC) proceeds to its High Luminosity (HL-LHC) phase, the Higgs physics programme continues to mature with the focus gradually shifting towards processes that probe the scalar sector beyond the single-Higgs level. The production of multiple Higgs bosons plays a pivotal role as it provides direct access to Higgs self-interactions and to possible additional scalar degrees of freedom associated with a richer scalar sector. Higgs pair production is a central component of the LHC Higgs programme from the point of view of Physics Beyond the Standard Model (BSM)~\cite{ATLAS:2023qzf,CMS:2022hgz}, while both theoretical and experimental efforts are increasingly exploring the prospects of triple Higgs production~\cite{Chen:2015gva,Papaefstathiou:2015paa,Papaefstathiou:2019ofh,Papaefstathiou:2020lyp,Papaefstathiou:2023uum,Karkout:2024ojx,Stylianou:2023tgg,Fuks:2017zkg,Kilian:2017nio,Lane:2024vur,Papaefstathiou:2025meh,ATLAS:2024xcs,Anisha:2024ljc,Anisha:2024ryj,Biermann:2024oyy,CMS:2025gos,Chiang:2025ecn,Dong:2025lkm,ATLAS:2025eii,ATLAS:2025cae,Abouabid:2024gms,CMS:2025jkb,CMS:2025ngq}, despite its extremely small rate in the Standard Model (SM)~\cite{deFlorian:2019app}. 

Triple Higgs production represents one of the most complex Higgs final states that can be contemplated at hadron colliders. Even when restricting attention to gluon fusion, the process involves a large number of loop-induced diagrams, multiple Higgs self-interactions, and a wide kinematic phase space. In the SM, this complexity is paired with an extremely small production rate, rendering the process effectively unobservable at the (HL-)LHC~\cite{deFlorian:2019app,Biermann:2024oyy,ATLAS:2025eii,ATLAS:2025cae}. However, in models with extended scalar sectors these rates can be largely enhanced, see e.g.~\cite{Karkout:2024ojx,Biermann:2024oyy,Abouabid:2024gms,Papaefstathiou:2020lyp,Ding:2026qto,Robens:2025nev,Samarakoon:2025iaq}. Therefore, inspired by \cite{Robens:2019kga,Papaefstathiou:2020lyp} (see also \cite{Abouabid:2024gms}), the experiments at the LHC have started to investigate triple Higgs final states. 
They have begun to place first direct constraints on triple Higgs production using Run~2 and early Run~3 data~\cite{ATLAS:2025eii,ATLAS:2025cae,CMS:2025gos}, typically targetting high-multiplicity final states such as $6b$, $4b2\tau$ and $4b2\gamma$ signatures, where the combinatorial background is particularly challenging. There is still no evidence for a signal within the SM, but bounds were imposed on a new physics scenario proposed in \cite{Robens:2019kga,Papaefstathiou:2020lyp}, as well as a simplified resonance enhanced scenario. The same results were also interpreted in terms of Higgs self-coupling modifiers.
The resulting limits on the SM cross-section (the present upper limit for the cross-section of gluon fusion triple Higgs production is $\sim 59~\text{fb}$~\cite{ATLAS:2024xcs}) still remain several orders of magnitude above the  prediction ($O(50-100~\text{ab})$ at $\sqrt{s} = 14~\text{TeV}$~\cite{deFlorian:2019app}). As a result, triple Higgs production is often discussed primarily in the context of future collider facilities or as a theoretical benchmark for Higgs self-interactions (see for example Refs.~\cite{Papaefstathiou:2015paa,Chen:2015gva,Dicus:2016rpf,Kilian:2017nio,Fuks:2017zkg,Agrawal:2017cbs,Liu:2018peg,Kilian:2018bhs,Chang:2019vez,Papaefstathiou:2019ofh,Abdughani:2020xfo,Papaefstathiou:2020lyp,Chiesa:2020awd,Haisch:2021hvy,Chiesa:2021qpr,Dermisek:2021mhi,Stylianou:2023tgg,Papaefstathiou:2023uum,Abouabid:2024gms,Biermann:2024oyy,ATLAS:2024xcs,Fuks:2025gjv,CMS:2025gos,ATLAS:2025eii,ATLAS:2025cae,Dong:2025lkm, Chiang:2025ecn}).

From a more model-independent perspective, triple Higgs production also provides a unique handle to discriminate between different effective field theory (EFT) descriptions of the scalar sector. In the Standard Model Effective Field Theory (SMEFT)~\cite{Weinberg:1978kz,Brivio:2017vri,Grzadkowski:2010es,Aebischer:2025qhh}, modifications to Higgs self-interactions are tightly correlated with deviations in single- and di-Higgs processes. Within the Higgs Effective Field Theory (HEFT)~\cite{Alonso:2015fsp,Buchalla:2012qq,Buchalla:2013rka,Buchalla:2017jlu,Brivio:2016fzo,Herrero:2021iqt,Herrero:2022krh,Alonso:2016oah,Cohen:2020xca,Sun:2022ssa,Brivio:2025yrr} however, such correlations can be significantly relaxed. As a result, triple Higgs final states can serve as a sensitive probe of extended scalar sectors even in scenarios where single and double Higgs observables remain close to their SM expectations~\cite{Anisha:2024ljc,Anisha:2024ryj}. The observation of anomalous features in triple Higgs production, would therefore provide a strong indication of non-linear dynamics in the scalar sector. Increasingly precise constraints can thus play an important role in distinguishing between SMEFTy and HEFTy new physics scenarios~\cite{Stylianou:2023tgg,Papaefstathiou:2023uum,Anisha:2024ljc,Anisha:2024ryj,Asiain:2026sio,Domenech:2025gmn,Brivio:2025sib,Fuks:2025gjv,Haisch:2025pql}.

Additionally, triple Higgs production is sensitive to the structure of the scalar potential and hence the nature of electroweak phase transitions; extensions of the SM scalar sector that allow for strong first-order phase transition typically induce sizeable modifications of Higgs self-interactions. While di-Higgs production probes mainly the trilinear coupling, triple Higgs production can provide complementary access to higher-order interactions offering an additional handle on the dynamics of electroweak symmetry breaking (EWSB) and baryogenesis-motivated scenarios~\cite{Karkout:2024ojx,Biermann:2024oyy}.

With the complicated final state in mind, some of the existing literature has focussed on simplified strategies to enhance the observability of resonant triple Higgs production (see e.g. \cite{Papaefstathiou:2025meh} for recent work). Such approaches often isolate specific topologies that can dominate when kinematically accessible, thereby providing a tractable framework for both theory predictions and experimental searches. While this strategy can capture the leading features in certain regions of parameter space, it does not fully reflect the complexity of the underlying dynamics. Away from strict resonance regimes, multiple interfering contributions, offshell effects, and non-trivial kinematic configurations can play an important role in shaping both rates and distributions. A more complete understanding of triple Higgs production therefore requires going beyond such simplified pictures and analysing the full kinematic structure of the process within realistic extended scalar sectors. 

In this work, we adopt this broader perspective and perform a detailed study of triple Higgs production in the next-to-minimal Two-Higgs-Doublet Model (N2HDM)~\cite{Muhlleitner:2016mzt} using it as a representative framework for an extended scalar sector. Our focus primarily lies on the kinematic properties of the final state and their interplay with the underlying scalar spectrum, their masses and couplings. 

This paper is organised as follows: in Sec.~\ref{sec:models} we swiftly introduce the N2HDM scalar sector under consideration, discussing the spectra involved in the production of triple Higgs final states from gluon-gluon fusion in hadron machines. We then describe our methodology in Sec.~\ref{sec:methodology}, including the details of parameter scans and event generation. We highlight the procedural aspects of selecting particular subprocesses involved in the full process. Sec.~\ref{sec:results} highlights the results of our analyses and our inference of the same from the point of view of numerical values, as well as kinematic aspects. We summarise and conclude in Sec.~\ref{sec:conc}.

%%%%%%%%%%%%%%%%%%%%%%%%%%%%%%%%%%
\section{Triple Higgs in the N2HDM}
\label{sec:models}
%%%%%%%%%%%%%%%%%%%%%%%%%%%%%%%%%%

In this work, we consider a specific extension to the SM scalar sector that gives rise to resonant contributions to triple Higgs production in gluon fusion, 
\begin{equation}
    g g \to h h h,
\end{equation}
with a particular emphasis on scenarios in which more than one heavy scalar can be produced onshell. In such cases, the amplitude receives contributions from multiple resonant propagators as well as from non-resonant diagrams and their mutual interference. The resulting cross-section and kinematic distributions cannot, in general, be understood in terms of a single factorised decay chain, even when restricting to regions of phase space that are naively expected to be dominated by resonant topologies~\cite{Abouabid:2024gms}.

In the SM, triple Higgs production proceeds entirely through continuum diagrams and is strongly suppressed by the smallness of the Higgs self-couplings and loop-induced nature of the process~\cite{deFlorian:2019app}. Extended Higgs sectors, on the other hand, can modify such interactions, as well as introduce additional scalar states that couple to gluons via heavy-quark loops. Schematically, one can write the partonic amplitude as,
\begin{equation}
    \mathcal{M} (gg \to hhh) = \mathcal{M}_{\text{non-resonant}} + \sum_i \mathcal{M}^{(i)}_{\text{resonant}},
\end{equation}
where $\mathcal{M}_\text{non-resonant}$ contains the SM continuum contributions and offshell Higgs exchange, while $\mathcal{M}_{\text{resonant}}^{(i)}$ denotes amplitudes mediated by intermediate scalar resonances $H_i$. Interference between these terms can play an essential role and significantly alter both the total rate and the shape of differential observables.

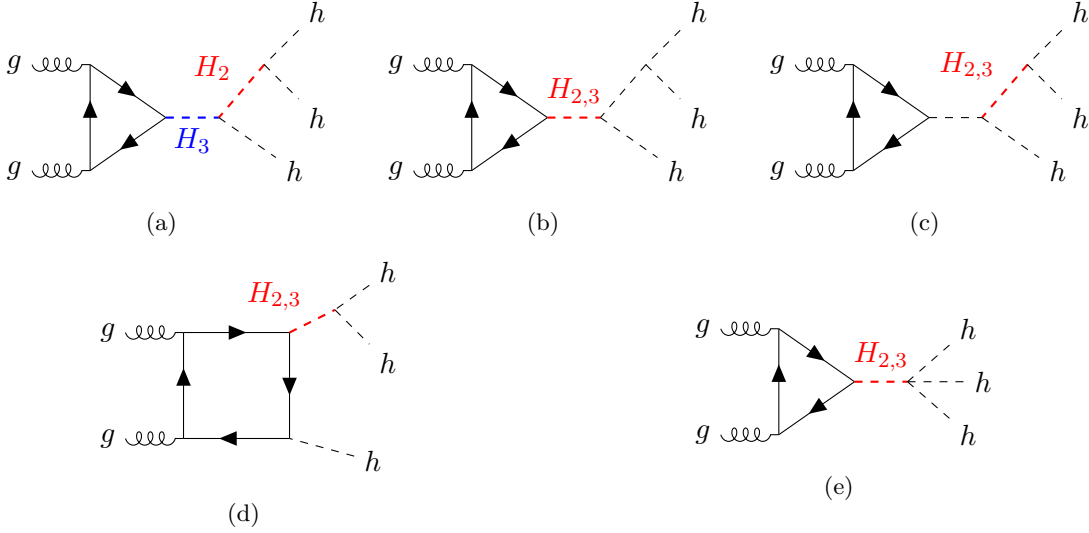
\begin{figure}[!t]
    \centering
    \hfill
    \begin{minipage}{0.33\textwidth}
    \centering
    \subfigure[]{\begin{tikzpicture}
    \begin{feynman}
        \vertex (g1) at (-1, 0.7) {$\large g$};
        \vertex (g2) at (-1, -0.7) {$\large g$};
        \vertex (a) at (0, 0.7);
        \vertex (b) at (0.,-0.7);
        \vertex (c) at (1, 0);
        \vertex (d) at (1.7, 0);
        \vertex (h) at (2.3,0.7);
        \vertex (h1) at (3, 1.4) {$\large h$};
        \vertex (h2) at (3, 0) {$\large h$};
        \vertex (h3) at (2.7,-0.7) {$\large h$};
        \diagram* {
            (g1) -- [gluon] (a),
            (g2) -- [gluon] (b),
            (a) -- [anti fermion] (b),
            (a) -- [fermion] (c) -- [fermion] (b),
            (c) -- [scalar, thick, blue, edge label' = $\color{blue}\large H_3\color{black}$] (d),
            (h) -- [scalar, thick, red, edge label' = $\color{red}\large H_2\color{black}$] (d) -- [scalar] (h3),
            (h1) -- [scalar] (h) -- [scalar] (h2),
        };
    \end{feynman}
    \end{tikzpicture}\label{fig:feyndiag1}}
    \end{minipage}\hfill\begin{minipage}{0.33\textwidth}
    \centering
    \subfigure[]{\begin{tikzpicture}
    \begin{feynman}
        \vertex (g1) at (-1, 0.7) {$\large g$};
        \vertex (g2) at (-1, -0.7) {$\large g$};
        \vertex (a) at (0, 0.7);
        \vertex (b) at (0.,-0.7);
        \vertex (c) at (1, 0);
        \vertex (d) at (1.7, 0);
        \vertex (h) at (2.3,0.7);
        \vertex (h1) at (3, 1.4) {$\large h$};
        \vertex (h2) at (3, 0) {$\large h$};
        \vertex (h3) at (2.7,-0.7) {$\large h$};
        \diagram* {
            (g1) -- [gluon] (a),
            (g2) -- [gluon] (b),
            (a) -- [anti fermion] (b),
            (a) -- [fermion] (c) -- [fermion] (b),
            (c) -- [scalar, thick, red, edge label = $\color{red}\large H_{2,3}\color{black}$] (d),
            (h) -- [scalar] (d) -- [scalar] (h3),
            (h1) -- [scalar] (h) -- [scalar] (h2),
        };
    \end{feynman}
    \end{tikzpicture}\label{fig:feyndiag2}}
    \end{minipage}\hfill\begin{minipage}{0.33\textwidth}
    \centering
    \subfigure[]{\begin{tikzpicture}
    \begin{feynman}
        \vertex (g1) at (-1, 0.7) {$\large g$};
        \vertex (g2) at (-1, -0.7) {$\large g$};
        \vertex (a) at (0, 0.7);
        \vertex (b) at (0.,-0.7);
        \vertex (c) at (1, 0);
        \vertex (d) at (1.7, 0);
        \vertex (h) at (2.3,0.7);
        \vertex (h1) at (3, 1.4) {$\large h$};
        \vertex (h2) at (3, 0) {$\large h$};
        \vertex (h3) at (2.7,-0.7) {$\large h$};
        \diagram* {
            (g1) -- [gluon] (a),
            (g2) -- [gluon] (b),
            (a) -- [anti fermion] (b),
            (a) -- [fermion] (c) -- [fermion] (b),
            (c) -- [scalar] (d),
            (h) -- [scalar, thick, red, edge label' = $\color{red}\large H_{2,3}\color{black}$] (d) -- [scalar] (h3),
            (h1) -- [scalar] (h) -- [scalar] (h2),
        };
    \end{feynman}
    \end{tikzpicture}\label{fig:feyndiag3}}
    \end{minipage}\hfill
    \begin{minipage}{0.48\textwidth}
    \centering
    \subfigure[]{\begin{tikzpicture}
    \begin{feynman}
        \vertex (g1) at (-1, 0.7) {$\large g$};
        \vertex (g2) at (-1, -0.7) {$\large g$};
        \vertex (a) at (0, 0.7);
        \vertex (b) at (0.,-0.7);
        \vertex (c) at (1.4, 0.7);
        \vertex (d) at (1.4, -0.7);
        \vertex (H) at (2, 1);
        \vertex (h1) at (2.5, -1) {$\large h$};
        \vertex (h2) at (2.7, 1.5) {$\large h$};
        \vertex (h3) at (2.7, 0.3) {$\large h$};
        \diagram* {
            (g1) -- [gluon] (a),
            (g2) -- [gluon] (b),
            (a) -- [anti fermion] (b) -- [anti fermion] (d),
            (a) -- [fermion] (c) -- [fermion] (d),
            (h1) -- [scalar] (d), 
            (c) -- [scalar,thick,red, edge label = $\large \color{red} H_{2,3} \color{black}$] (H),
            (h2) -- [scalar] (H) -- [scalar] (h3),
        };
    \end{feynman}
    \end{tikzpicture}\label{fig:feyndiag4}}\\
    \end{minipage}\hfill\begin{minipage}{0.48\textwidth}
    \centering
    \vspace{0.4cm}
    \subfigure[]{\begin{tikzpicture}
    \begin{feynman}
        \vertex (g1) at (-1, 0.7) {$\large g$};
        \vertex (g2) at (-1, -0.7) {$\large g$};
        \vertex (a) at (0, 0.7);
        \vertex (b) at (0.,-0.7);
        \vertex (c) at (1, 0);
        \vertex (d) at (1.7, 0);
        \vertex (h1) at (2.5, 0.7) {$\large h$};The wall’s longitudinal momentum
        \vertex (h2) at (2.5, -0.7) {$\large h$};
        \vertex (h3) at (2.7,0) {$\large h$};
        \diagram* {
            (g1) -- [gluon] (a),
            (g2) -- [gluon] (b),
            (a) -- [anti fermion] (b),
            (a) -- [fermion] (c) -- [fermion] (b),
            (c) -- [scalar, thick, red, edge label = $\color{red}\large H_{2,3}\color{black}$] (d),
            (h1) -- [scalar] (d) -- [scalar] (h2),
            (d) -- [scalar] (h3),
        };
    \end{feynman}
    \end{tikzpicture}\label{fig:feyndiag5}}\\ 
    \end{minipage}
    
    \caption{Representative Feynman topologies relevant for the resonant production of three Higgs bosons with 2 additional physical scalar degrees of freedom. %{\bf tr need to discuss more}
    }
    \label{fig:feyndiags}
    \end{figure}

A characteristic rich realisation of this structure arises in Higgs sectors featuring three neutral scalars that mix with the observed Higgs boson ($H_1=h$, with $m_h = 125~\text{GeV}$). If the scalar spectrum exhibits a hierarchical pattern, 
\begin{equation}
    m_{H_3} > m_{H_2} + m_{h},~\text{and}~m_{H_2} > 2 m_h,
\end{equation}
the process, 
\begin{equation}
    g g \to H_3 \to H_2 h \to h h h 
\end{equation}
can proceed through two successive onshell decays, giving rise to a double-resonant contribution to the $g g \to h h h $ amplitude (\textit{cf.} Fig.~\ref{fig:feyndiag1}). This configuration characterised by two distinct invariant mass scales can leave pronounced imprints in kinematic observables such as the di- and tri-Higgs invariant masses, and the Higgs transverse-momentum spectra. Furthermore, the couplings relevant in this process are the Yukawa-like interactions with the heaviest scalar $H_3$ ($y_{t\bar t H_3},~y_{b \bar b H_3}$) and the purely trilinear scalar couplings between $H_3,H_2,h$ ($\lambda_{123}$), and $H_2$ and two SM Higgses ($\lambda_{112}$). Such double-resonant topologies are often used as a simplified description of the process~\cite{Papaefstathiou:2025meh}, effectively factorising it into sequential onshell decays. While this approximation may capture the dominant features in regions with well-separated mass scales, it does not fully account for the interplay among various topologies, and interference effects. At the same time, these resonant contributions coexist with diagrams in which one or both intermediate states are offshell, making it non-trivial to completely isolate a purely resonant signal. The impact of such effects in the presence of more than 2 scalar resonances has been previously investigated in the context of the two real singlet extension of the SM (TRSM), where it has been shown that even in the presence of well-separated mass eigenstates and pronounced resonant enhancements, the full $g g \to h h h$ process can deviate significantly from simplified descriptions based on sequential onshell decays~\cite{Papaefstathiou:2020lyp,Abouabid:2024gms}. These findings motivate the question of whether similar conclusions persist in other structured Higgs-sector extensions, where additional correlations between masses and couplings arise and where the scalar spectrum is tied to the existing Higgs data.
%%%%%%%%%%%%%%%%%%%%%%%%%%%%%%%%%%%%%%%%%%%%%%%%%%%%%%%%%%
%%%%%%%%%%%%%%%%%%%%%%%%%%%%%%%%%%%%%%%%%%%%%%%%%%%%%%%%%%
Motivated by this, one can focus on extensions of the two-Higgs-doublet model (2HDM) that naturally realise $\geq3$ mixed neutral scalars: the complex 2HDM (C2HDM)~\cite{Fontes:2017zfn}, and the N2HDM~\cite{Muhlleitner:2016mzt}. In both cases, the scalar spectrum contains three neutral mass eigenstates that can participate in resonant Higgs cascades while remaining consistent with current constraints from Higgs signal-strength data and direct searches for additional scalars. 
The scalar potential of the 2HDM is given by~\cite{Branco:2011iw},
\begin{equation}
    \label{eq:r2dm}
    \begin{split}
    V_{\text{2HDM}}&=m_{11}^{2}\Phi_{1}^{\dagger}\Phi_{1}+m_{22}^{2}\Phi_{2}^{\dagger}\Phi_{2} - \left( m_{12}^{2}\Phi_{1}^{\dagger}\Phi_{2}+{\text{h.c.}} \right)+\frac{\lambda_{1}}{2}\left(\Phi_{1}^{\dagger}\Phi_{1}\right)^{2}+\frac{\lambda_{2}}{2}\left(\Phi_{2}^{\dagger}\Phi_{2}\right)^{2}\\
    &+\lambda_{3}\left(\Phi_{1}^{\dagger}\Phi_{1}\right)\left(\Phi_{2}^{\dagger}\Phi_{2}\right)+\lambda_{4}\left(\Phi_{1}^{\dagger}\Phi_{2}\right)\left(\Phi_{2}^{\dagger}\Phi_{1}\right) + \left[\frac{\lambda_{5}}{2}\left(\Phi_{1}^{\dagger}\Phi_{2}\right)^{2}+\text{h.c.}\right],
    \end{split}
\end{equation}
with a softly-broken $\mathbb{Z}_2$ symmetry, where the doublets 
\begin{equation}
    \Phi_i = \frac{1}{\sqrt{2}} \begin{pmatrix}
        \phi_i^+ \\
        v_i + \varphi_i + i a_i
    \end{pmatrix},\quad i = 1,2,
\end{equation}
transform as $(\mathbf{1},\mathbf{2},1/2)$ under the SM gauge symmetry. In the C2HDM, the parameters $m_{12}^2$ and $\lambda_5$ have different complex phases (so that they cannot be removed by a field redefinition), leading to explicit CP-violation in the scalar sector. 

If all the parameters in $V_{\text{2HDM}}$ are real, and the scalar sector is instead extended by a real gauge-singlet field $\Phi_S$, 
\begin{equation}
    \Phi_S = v_S+\varphi_3,
\end{equation}
then one obtains the N2HDM, with additional terms in the potential,
\begin{equation}
    V_{\text{N2HDM}} = V_{\text{2HDM}} + \frac{1}{2} m_S^2 \Phi^2_S + \frac{\lambda_6}{8} \Phi_S^4 + \frac{\lambda_7}{2} (\Phi^\dagger_1\Phi_1) \Phi_S^2 + \frac{\lambda_8}{2} (\Phi^\dagger_2\Phi_2) \Phi_S^2.
\end{equation}
The N2HDM has in total 12 independent real parameters~\cite{Muhlleitner:2016mzt} which we will discuss more in the following. The presence of the CP-admixture in the case of the C2HDM, and the additional singlet scalar in the case of the N2HDM relaxes the coupling correlations of the minimal 2HDM and allows for a wider range of scalar mass hierarchies relevant for multi-resonant Higgs production~\cite{Biermann:2024oyy}. Since stringent bounds from electric dipole moments allow only a small CP admixture in the C2HDM~\cite{Muhlleitner:2017dkd}, the resulting modifications to di-Higgs and triple Higgs production are typically strongly correlated. Since the N2HDM provides greater freedom in the scalar sector, we choose to perform our analysis within the N2HDM framework.

After EWSB, the two doublet fields acquire real vacuum expectation values (vevs) $v_1$, $v_2$, and the singlet field acquires a real vev $v_S$. The CP-even neutral fields $(\varphi_1,\varphi_2,\varphi_3)$ mix through a $3\times 3$ mass matrix obtained from the second derivatives of the scalar potential evaluated at the minimum,
\begin{equation}
M^2_{\text{scalar}} =
\begin{pmatrix}
\lambda_1 c_{\beta}^2 v^2 + t_{\beta} m_{12^2} & \lambda_{345} c_{\beta} s_\beta v^2 - m_{12}^2 & \lambda_7 c_\beta v v_S \\
\lambda_{345} c_{\beta} s_\beta v^2 - m_{12}^2 & \lambda_2 s_{\beta}^2 v^2 + m_{12^2}/t_\beta & \lambda_8 s_\beta v v_S \\
\lambda_7 c_\beta v v_S  & \lambda_8 s_\beta v v_S & \lambda_6 v_S^2
\end{pmatrix},
\end{equation}
where, $\lambda_{345} = \lambda_3+\lambda_4+\lambda_5$, $t_\beta \equiv \tan\beta = v_2/v_1$, $s_\beta\equiv \sin \beta$, $c_\beta \equiv \cos\beta$, and the total electroweak vev $v=\sqrt{v_1^2+v_2^2}$. The physical CP-even mass eigenstates $H_i$ are obtained by diagonalising this matrix with an orthogonal rotation,
\begin{equation}
    H_i = R_{ij} \varphi_j, \qquad 
    R\, M^2_{\text{scalar}}\, R^T = \mathrm{diag}(m^2_{H_1}, m^2_{H_2}, m^2_{H_3}).
\end{equation}
The rotation matrix $R$ can be parametrised in terms of three mixing angles $\alpha_1$, $\alpha_2$, and $\alpha_3$ as
\begin{equation}
R =
\begin{pmatrix}
c_1 c_2 & s_1 c_2 & s_2 \\
-(c_1 s_2 s_3 + s_1 c_3) & c_1 c_3 - s_1 s_2 s_3 & c_2 s_3 \\
- c_1 s_2 c_3 + s_1 s_3 & -(c_1 s_3 + s_1 s_2 c_3) & c_2 c_3
\end{pmatrix},
\end{equation}
where $s_i \equiv \sin\alpha_i$ and $c_i \equiv \cos\alpha_i$.

The scalar sector of the N2HDM can be conveniently expressed in terms of physical parameters after EWSB. A suitable choice comprises
\begin{equation}
    m_{H_1},~m_{H_2},~m_{H_3},~m_A,~m_{H^\pm},~\alpha_1,~\alpha_2,~\alpha_3,~m_{12}^2,~\tan{\beta},~v_S,~v,
    \label{eqn:n2hdmpars}
\end{equation}
where, $m_{H_1}, m_{H_2}, m_{H_3}$ are the three CP-even scalar masses, $m_A$ is the CP-odd mass, and $m_{H^\pm}$ is the charged Higgs mass. Among the given parameters, the total vev ($v$) and $m_{H_1}$ are fixed to the electroweak vev ($v_{\text{EW}} \simeq 246~\text{GeV}$), and the mass of the detected Higgs boson ($m_{h} \simeq 125~\text{GeV}$) from experiments. This set of parameters provides a one-to-one mapping on to the parameters of the scalar potential. The parameters are, however, subject to theoretical consistency conditions, such as perturbative unitarity, boundedness of the potential from below, and the requirement that the electroweak vacuum corresponds to the global minimum. In addition, agreement with electroweak precision data and collider measurements further restricts the allowed parameter space. See e.g. \cite{Muhlleitner:2016mzt} for more details as well as \cite{Wittbrodt:2016fuk,Wittbrodt:2019bsu} for a more thorough discussion.

For the numerical analysis, it is convenient to adopt the input parameterisation used in \texttt{ScannerS}~\cite{Coimbra:2013qq, Muhlleitner:2020wwk, Muhlleitner:2016mzt}, where the scalar mixing is traded for quantities more directly related to Higgs couplings. In this basis, the parameter set is given by 
\begin{equation}
m_{H_1},~m_{H_2},~m_{H_3},~m_A,~m_{H^\pm},~\tan\beta,~c_{hVV}^2,~c_{ht\bar t}^2,~\text{sign}(R_{13}),~R_{23},~m_{12}^2,~v_S,~v. 
\label{eqn:n2hdmparsnew}
\end{equation}
Here, $c_{hVV}^2$ and $c_{htt}^2$ denote the squared couplings of the SM-like Higgs boson to vector bosons and top quarks, normalised to their SM values. The parameters $R_{13},~R_{23}$ describe the mixing between $H_{1,2}$ and the singlet field, respectively. Note that already the parametrisation in Eq.~(\ref{eqn:n2hdmpars}) contains a redundancy in the definition of the mixing angles: if all possible ranges were considered, we would have regions with different choices that lead to the same physical predictions (see e.g. \cite{Ferreira:2014naa} for a more thorough discussion on this in the context of the 2HDM)\footnote{We thank Rui Santos for useful discussions regarding this point.}. In order to not scan over such redundant regions in the parameter space, the scan ranges were fixed in \texttt{ScannerS} using $\text{sign}(R_{13})$. The second sign is fixed by the assumption 
\begin{equation}
    c_{hVV}\times c_{h t \bar t} > 0.
\end{equation}
The choice in Eq.~(\ref{eqn:n2hdmparsnew}) enables us to choose input parameters in such a way that points generated in the scan are efficiently aligned with experimentally relevant observables.

In the following, the lightest neutral CP-even states is identified with the observed Higgs boson at $125~\text{GeV}$, while the remaining free parameters are varied to explore the resonant production of three Higgs bosons, where in particular the couplings $\lambda_{123}$ and $\lambda_{112}$ are of importance. We now provide the details of the parameter scan and the implementation of the various constraints.

%%%%%%%%%%%%%%%%%%%%%%%%%%%%%%%%%%
\section{Methodology}
\label{sec:methodology}
%%%%%%%%%%%%%%%%%%%%%%%%%%%%%%%%%%
We shall first describe the methodology of event-generation for this work. 
As a first step, we scan over the parameter space of the selected model using the publicly available code \texttt{ScannerS}~\cite{Coimbra:2013qq, Muhlleitner:2020wwk, Muhlleitner:2016mzt}. 
Experimental bounds from searches for new scalar particles and measurements of the 125~GeV Higgs boson at colliders are applied via the public code \texttt{HiggsTools}~\cite{Bahl:2022igd} which is a new implementation of the former \texttt{HiggsBounds}~\cite{Bechtle:2008jh, Bechtle:2011sb, Bechtle:2012lvg, Bechtle:2013wla, Bechtle:2015pma, Bechtle:2020pkv, Bahl:2021yhk} and \texttt{HiggsSignals}~\cite{Bechtle:2013xfa, Stal:2013hwa, Bechtle:2014ewa, Bechtle:2020uwn}. Along with checking experimental bounds via \texttt{HiggsTools}, $\texttt{ScannerS}$ also checks bounds from electroweak precision observables and theoretical bounds from unitarity, boundedness-from-below and vacuum stability~\cite{Hollik:2018wrr, Ferreira:2019iqb}. The ranges of the input parameters that are kept fixed throughout the scans are shown in Tab.~\ref{tab:scanners}. The neutral scalar masses $m_{H_1}, m_{H_2}, m_{H_3}$ are not included in this table, as they are treated separately: depending on the analysis, they are either scanned over or fixed to specific values in order to realise particular kinematic configurations of interest.
\begin{table}[!t]
\centering
\begin{tabular}{|lccccccccc|}
\hline
 &  $m_A$ & $m_{H^\pm}$ & $\tan{\beta}$ & $c_{hVV}^2$ & $c_{htt}^2$ & sign($R_{13}$) & $R_{23}$ & $m_{12}^2$ & $v_S$ \\
\hline
min. & 30 & 580 & 0.8 & 0.9 & 0.8 & -1 & -1 & $10^{-3}$ & 1 \\
max. & 1500 & 1500 & 20 & 1.1 & 1.2 & 1 & 1 & $5\times10^5$  & 3000 \\
\hline
\end{tabular}
\caption{Parameter ranges as \texttt{ScannerS} input. \label{tab:scanners}
}
\end{table}

From the points allowed by the constraints mentioned above, we select the points where the cross-section of the double-resonance process is maximal. This can be obtained from the \texttt{ScannerS+HiggsTools} output by constructing the factorised double-resonant contribution:
\begin{equation}
\sigma_{\text{DR}} = \sigma (g g \to H_3) \times \text{BR} (H_3 \to H_2 h) \times \text{BR} (H_2 \to h h) \, .
\end{equation}
(A `maximal' double-resonance cross-section amounts to about $\sigma_{\text{DR}} \sim O(60~\text{fb})$, chosen to remain close to the upper bound on the total SM triple-Higgs production cross-section of 59~fb~\cite{ATLAS:2024xcs}\footnote{At present, no dedicated upper limits exist for resonant triple-Higgs production in the N2HDM. While larger cross-sections are phenomenologically interesting, such enhancements may also manifest in the di-Higgs channel due to correlations induced by the fixed SM gauge charges of the scalar sector. The bound adopted here is also motivated by the mass-dependent limits on the triple-Higgs cross-section obtained in the TRSM, as shown in Ref.~\cite{ATLAS:2024xcs}.})
{We additionally select a set of parameter points for which the double-resonant cross section is significantly lower, thereby illustrating different phenomenological scenarios. The scans performed using \texttt{ScannerS} employs state-of-the-art higher-order inputs: production cross sections are evaluated at next-to-next-to-leading order accuracy with \texttt{SuSHI}~\cite{Harlander:2012pb,Harlander:2016hcx}, while decay widths and branching ratios, including higher-order corrections, are computed with \texttt{N2HDecay}~\cite{Engeln:2018mbg}.} 

We generate our events for the selected points using the Monte Carlo event generator \texttt{MadGraph5\_aMC@NLO}~\cite{Alwall:2014hca, Frederix:2018nkq} at one-loop level in QCD, including both top and bottom quark contributions. We use the \texttt{FeynRules}~\cite{Alloul:2013bka} implementation of the N2HDM scalar sector,  using \texttt{NLOCT}~\cite{Degrande:2014vpa}, interfaced to \texttt{MadGraph5} with a \texttt{UFO}~\cite{Degrande:2011ua,Darme:2023jdn} model file, developed previously for the analysis performed in Ref.~\cite{Biermann:2024oyy}. The events are generated for the full process (including all relevant topologies), only the double-resonant (DR) process (\textit{cf.} Fig.~\ref{fig:feyndiag1}), and all processes except the double-resonance (Non-DR) in order to assess the interference effects. We also generate events for processes involving the box diagrams shown on Fig.~\ref{fig:feyndiag4}: (1) involving $H_2$ (box-$H_2$) and (2) involving $H_3$ (box-$H_3$) in the respective propagators to further assess the rich peak structures in the kinematic distributions. 

In order to single out these different processes we use a combination of setting specific couplings and diagrams to zero in \texttt{MadGraph5} (the relevant settings are highlighted in Tab.~\ref{tab:madgraph_settings}).
\begin{table}[!t]
    \centering
    \begin{tabular}{|p{1.5cm}|p{12cm}|}
        \hline
        Process & Settings\\
        \hline 
        full & Default \texttt{MadGraph} command: \newline\texttt{generate p p > h h h [QCD] NP=3 QCD=2 QED=3}\newline All couplings set to their default values returned by \texttt{ScannerS}.\\
        \hline 
        DR & $\lambda_{112}, \lambda_{123}, y_{t\Bar{t}H_3}, y_{b\Bar{b}H_3} \neq 0 $, all other couplings set to 0.\\
        \hline 
        Non-DR & $\lambda_{123}=0$, all other couplings non-zero. \\
        \hline 
        box-$H_3$ & \vtop{\hbox{\strut $\lambda_{113}, y_{t\Bar{t}H_3}, y_{b\Bar{b}H_3}, y_{t\bar{t}h}, y_{b\Bar{b}h} \neq 0 $, all other couplings set to 0,}\hbox{\strut \texttt{MadGraph} coupling orders set to: \texttt{NP=3} \texttt{QCD=2} \texttt{QED=2},}\hbox{\strut diagrams of the type as in Fig.~\ref{fig:feyndiag3} with an $H_3$ propagator are set to 0}\hbox{\strut by hand}}
        \\
        \hline 
        box-$H_2$ & \vtop{\hbox{\strut $\lambda_{112}, y_{t\Bar{t}H_2}, y_{b\Bar{b}H_2}, y_{t\Bar{t}h}, y_{b\Bar{b}h} \neq 0$, all other couplings set to 0,}\hbox{\strut \texttt{MadGraph} coupling orders set to: \texttt{NP=3} \texttt{QCD=2} \texttt{QED=2},}\hbox{\strut diagrams of the type as in Fig.~\ref{fig:feyndiag3} with an $H_2$ propagator are set to 0}\hbox{\strut by hand}}
        \\
        \hline
    \end{tabular}
    \caption{
    The settings in \texttt{MadGraph} in order to single out the different processes. Here \texttt{QCD} refers to the strong coupling order (setting the number of gluon-gluon-fermion couplings), and \texttt{QED} sets the coupling order of the fermion-fermion-scalar couplings. The \texttt{NP} coupling order sets the number of pure scalar-scalar couplings.
    }
    \label{tab:madgraph_settings}
\end{table}
For the DR and Non-DR processes it is sufficient to identify the relevant couplings for that process and set all other couplings to zero. To isolate the box diagrams, box-$H_2$ and box-$H_3$, the easiest way is to use a combination of setting certain couplings to zero, setting the coupling orders to \texttt{QED=2} in order to exclude pentagon diagrams for the SM process, and finally, setting diagrams of the type as in Fig.~\ref{fig:feyndiag3} to zero by hand. Alternatively one could also go through all diagrams and set all diagrams which are not relevant to zero by hand. This allows us to distinguish between the contributions from the different kinds of topologies individually. 

We generate $10^4$ events for every point at a centre-of-mass energy of $\sqrt{s}=13.6$~GeV, using these to plot the number of expected events at the LHC using an integrated luminosity of $L_{\text{int.}} = 300~\text{fb}^{-1}$ for the following kinematic observables: the invariant masses $M_{hh}$ and $M_{hhh}$ of the di- and tri-SM-like Higgs boson system, respectively, and the transverse momentum $p_{T}$. The invariant masses are defined as:
\begin{equation}
\begin{split}
    M_{h_ih_j} &\equiv \sqrt{(p_{h_i} + p_{h_j})^2} , \\
    M_{hhh} &\equiv \sqrt{(p_{h_1} + p_{h_2} + p_{h_3})^2} ,
\end{split}
\end{equation}
where the subscripts $i,j=1,2,3$ label the individual SM-like Higgs bosons in the final state and all combinations are taken in the variable $M_{hh}$.

%%%%%%%%%%%%%%%%%%%%%%%%%%%%%%%%%%
\begin{figure}[!t]
    \centering
    \includegraphics[width=0.48\linewidth]{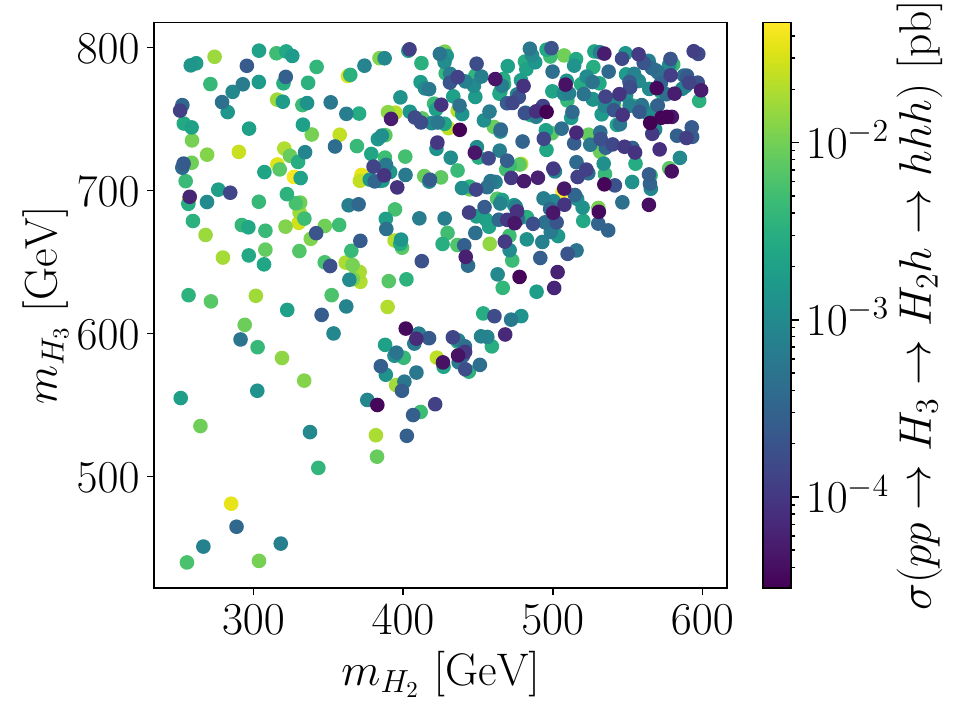}
    \includegraphics[width=0.48\linewidth]{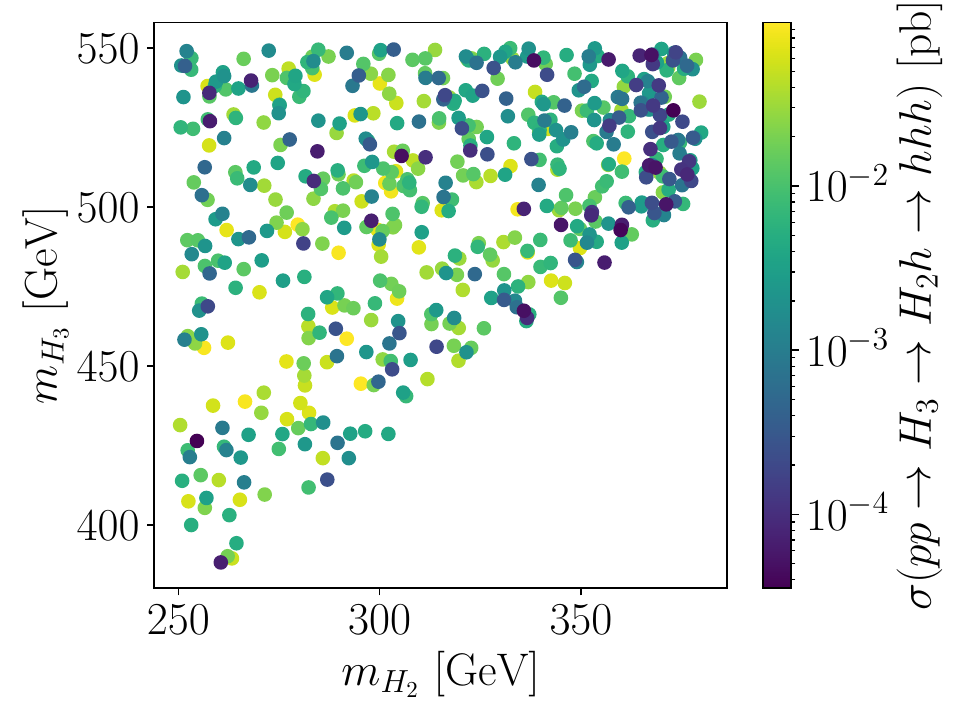}
    \caption{The dependence of the cross-section of the double-resonant process on the masses of the heavier neutral scalar degrees of freedom, obtained directly from \texttt{ScannerS}~\cite{Muhlleitner:2020wwk}. The plot on the right is obtained by zooming in to the mass-regimes where the double-resonance is numerically maximised, including additional scan points tailored to this region.
    }
    \label{fig:xsec_masses}
\end{figure}
%%%%%%%%%%%%%%%%%%%%%%%%%%%%%%%%%%
%%%%%%%%%%%%%%%%%%%%%%%%%%%%%%%%%%
\section{Results}
\label{sec:results}
%%%%%%%%%%%%%%%%%%%%%%%%%%%%%%%%%%
We now turn to our results. The full process $p p \to h h h$ receives contributions from a large number of topologies which interfere with one another. As a result, it is not straightforward to provide a fully transparent assessment of resonant enhancements of the cross-section, particularly in the presence of two additional scalar degrees of freedom. Nevertheless, as pointed out in Refs.~\cite{Biermann:2024oyy,Papaefstathiou:2020lyp}, the di- and tri-Higgs invariant masses, $M_{hh}$ and $M_{hhh}$, can in principle provide some guidance in disentangling the dominant contributions entering the full process (we shall discuss these for individual points later in the section). For the $M_{hh}$ and $M_{hhh}$ distributions, peaks will usually show around the regions of $m_{H_2}$ and $m_{H_3}$ depending on the exact kinematic configurations. 
In the $M_{hh}$ distributions, peaks usually indicate a two-body decay from a heavier scalar to two SM-like Higgs bosons, when kinematically allowed. For $M_{hhh}$ distributions, peaks usually indicate a decay from a heavier scalar into three SM-like Higgs bosons if they are kinematically feasible, as well as the full decay chain of the double-resonance (i.e., $H_3 \to H_2 h \to hhh$). 
One can additionally consider the azimuthal angle $\phi$ and pseudorapidity $\eta$ distributions, however, the differences between the various contributing topologies are comparatively negligible negligible (see App.~\ref{app:angular} for more details). 

As a first step, we examine the dependence of the cross-section of our target process on the masses of the heavier scalar states. 
For kinematically allowed parameter points, one may naively expect the cross-section to be maximised for $m_{H_3} \sim 3 m_h$ and $m_{H_2} \sim 2 m_h$~\cite{Papaefstathiou:2020lyp}. As illustrated in Fig.~\ref{fig:xsec_masses}, the bulk of parameter points yielding the largest cross-sections indeed populate this region of parameter space, with a number of outliers. 

On the basis of this behaviour, the following benchmark points are selected, corresponding to parameter configurations for which the double-resonance cross-section is maximised:\\

%%%%%%%%%%%%%%%%%%%%%%%%%%%%%%%%%%
\begin{minipage}{0.48\textwidth}
    \begin{itemize}
        \item $m_{H_2}, m_{H_3} = 254,~394~\text{GeV}$ 
        \item $m_{H_2}, m_{H_3} = 395,~754~\text{GeV}$ 
        \item $m_{H_2}, m_{H_3} = 400,~537~\text{GeV}$ 
        \item $m_{H_2}, m_{H_3} = 600,~907~\text{GeV}$ 
    \end{itemize}
\end{minipage}\hfill\begin{minipage}{0.48 \textwidth}
    \begin{itemize}
        \item $m_{H_2}, m_{H_3} = 269,~450~\text{GeV}$ 
        \item $m_{H_2}, m_{H_3} = 507,~700~\text{GeV}$ 
        \item $m_{H_2}, m_{H_3} = 400,~765~\text{GeV}$ 
        \item $m_{H_2}, m_{H_3} = 600,~732~\text{GeV}$ 
    \end{itemize}
\end{minipage}
%%%%%%%%%%%%%%%%%%%%%%%%%%%%%%%%%%
\\

We first aim to quantify how much of the total cross-section is accounted for by the double-resonance contribution by directly comparing the numerical values of the corresponding cross-sections. For each of the selected mass benchmarks, the scalar masses are fixed while the remaining parameters are scanned using the \texttt{ScannerS}+\texttt{HiggsTools} toolchain. From this scan, ten parameter points are selected which maximise the cross-section of the double-resonance process, subject to the condition $\sigma_{\text{DR}} \sim O(60~\text{fb})$.

\begin{figure}[!t]
\centering
\includegraphics[width=0.8\linewidth]{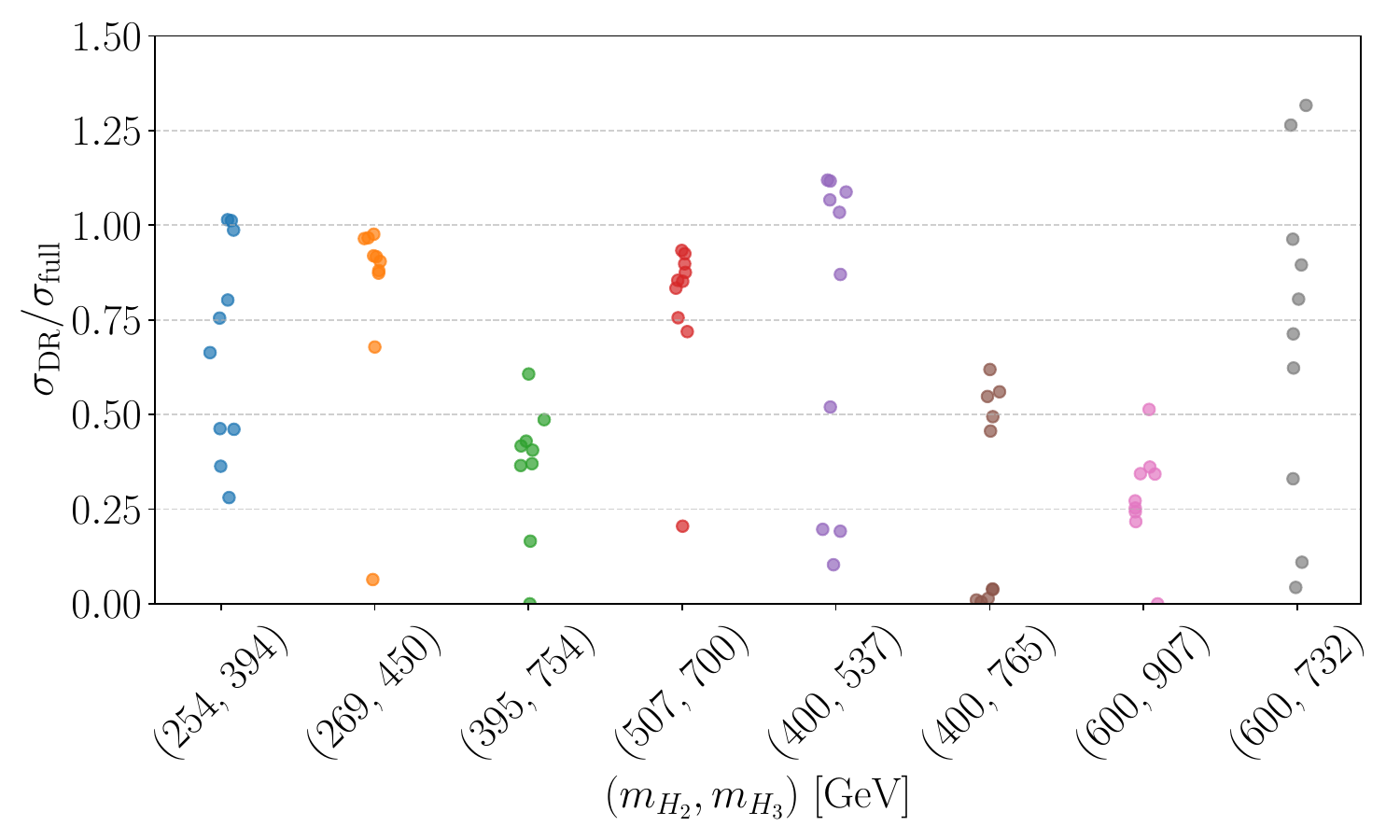}
\caption{Ratio of the double-resonant cross-sections ($\sigma_{\text{DR}}$) with respect to the cross-section from the full process ($\sigma_{\text{full}}$) for the various mass benchmarks under consideration}
\label{fig:xsecpercentages}
\end{figure}

For each of these parameter points, events are generated with \texttt{MadGraph5} for both the full $p p \to h h h $ process and the double-resonance topology in order to determine the corresponding integrated cross-sections. Fig.~\ref{fig:xsecpercentages} shows the ratio of the double-resonant cross-section to that of the full process for various benchmark scenarios. 

%%%%%%%%%%%%%%%%%%%%%%%%%%%%%%%%%%
%%%%%%%%%%%%%%%%%%%%%%%%%%%%%%%%%%
Several observations can be made at this point. First, apart from some outliers, the benchmarks with $M_{H_2} \sim 2 m_h$ and $M_{H_3} \sim 3 m_h$ show that a sizeable fraction of the total cross-section originates from the double-resonance topology, as expected from the kinematics of the target process. Second, even in cases where $M_{H_2}$ is significantly larger than $2 m_h$, if the relation $M_{H_3}$ lies in the threshold of $ \sim M_{H_2}+m_h$ allowing the decay $H_3 \to H_2 h$ to occur onshell, parameter points can still be found where the double-resonance contribution approaches the full cross-section, leading to a larger spread in the ratio. Third, scenarios with more widely separated mass scales exhibit both a smaller spread, and systematically lower values of this ratio. Finally, a few points yield ratios larger than unity, indicating the presence of destructive interference effects among the various topologies contributing to the full $p p \to h h h$ amplitude. This observation also highlights an important limitation of interpreting the ratio purely as a measure of the fraction of the total cross-section arising from the double-resonance contribution. Since interference effects are present at the amplitude level, they can in principle occur throughout the parameter space and are not restricted to particular mass configurations. Therefore, the numerical ratio of the double-resonance and full cross-sections should be regarded only as an indicative measure. 

Having examined the ratios at the level of cross-sections, one may further ask whether, for fixed mass configurations, the relative contribution of the double-resonance topology can be directly linked to the underlying scalar self-couplings. Naively, one might expect this fraction to be indicative of the product of $\lambda_{123}$ and $\lambda_{112}$ entering the subsequent decays, which does in fact decide the individual amplitude of the double-resonance process~\cite{Papaefstathiou:2025meh}. However, we observe no clear correlation of this product with the ratio, across the selected parameter points for any of the mass benchmarks. This further reflects that the ratio is sensitive to a non-trivial interplay of parameters and diagrams beyond the scalar couplings alone, including widths and interference effects. This further elucidates the importance of considering the full process rather than relying solely on the factorised double-resonance contribution. {To analyse this in further detail, we select representative benchmark scenarios, with different kinematic configurations (i.e., mass-spectra among the heavier scalars) where the the contribution of our target double-resonance process to the total cross-section is significant, as well as scenarios where even with the favoured kinematic configuration, the double-resonant process is subdominant. The masses, widths, and the coupling structures relevant for the considered benchmark points are tabulated in Tab.~\ref{tab:couplings}.}

%%%%%%%%%%%%%%%%%%%%%%%%%%%%%%%%%%
\begin{table}[!htpb]
\centering
\resizebox{\textwidth}{!}{
\begin{tabular}{|lcccccccc|}
\hline
 &  $m_{H_2}$~[GeV] & $\Gamma_{H_2}$~[GeV] & $m_{H_3}$~[GeV] & $\Gamma_{H_3}$~[GeV] & $y_{t \bar t H_3}$ & $y_{b \bar b H_3}$ & $\lambda_{112}$~[GeV] & $\lambda_{123}$~[GeV] \\
\hline
BP-1 & 254 & 0.43 & 394 & 2.57 & 0.973 & 0.927 & -169.8 & 103.3 \\
BP-2 & 254 & 0.18 & 394 & 3.25 & 0.972 & 0.992 & -97.5 & -98.1 \\
BP-3 & 400 & 1.98 & 537 & 7.08 & 1.003 & 0.883 & -519.9 & -646.1 \\
BP-4 & 400 & 1.26 & 537 & 8.00 & 1.007 & 0.921 & -419.0 & 716.0 \\
BP-5 & 400 & 0.80 & 537 & 4.53 & 0.996 & 0.955 & 499.0 & 55.5 \\
BP-6 & 600 & 6.92 & 732 & 17.97 & 1.009 & 0.915 & -618.3 & 460.8 \\
\hline
\end{tabular}
}
\caption{Masses, widths of the heavier resonances, and coupling structures relevant for the double-resonant process for the benchmark points considered in our analysis.}
\label{tab:couplings}
\end{table}
%%%%%%%%%%%%%%%%%%%%%%%%%%%%%%%%%%

%%%%%%%%%%%%%%%%%%%%%%%%%%%%%%%%%%
\subsection{Mass threshold maximising the double-resonance}
%%%%%%%%%%%%%%%%%%%%%%%%%%%%%%%%%%

We now examine a few representative scenarios in order to highlight the non-trivial kinematic features that can arise and contribute to the triple-Higgs production process. Our analysis is performed at the parton level, focusing on the final-state Higgs bosons. 

\begin{figure}[!t]
\centering
\includegraphics[width=0.48\linewidth]{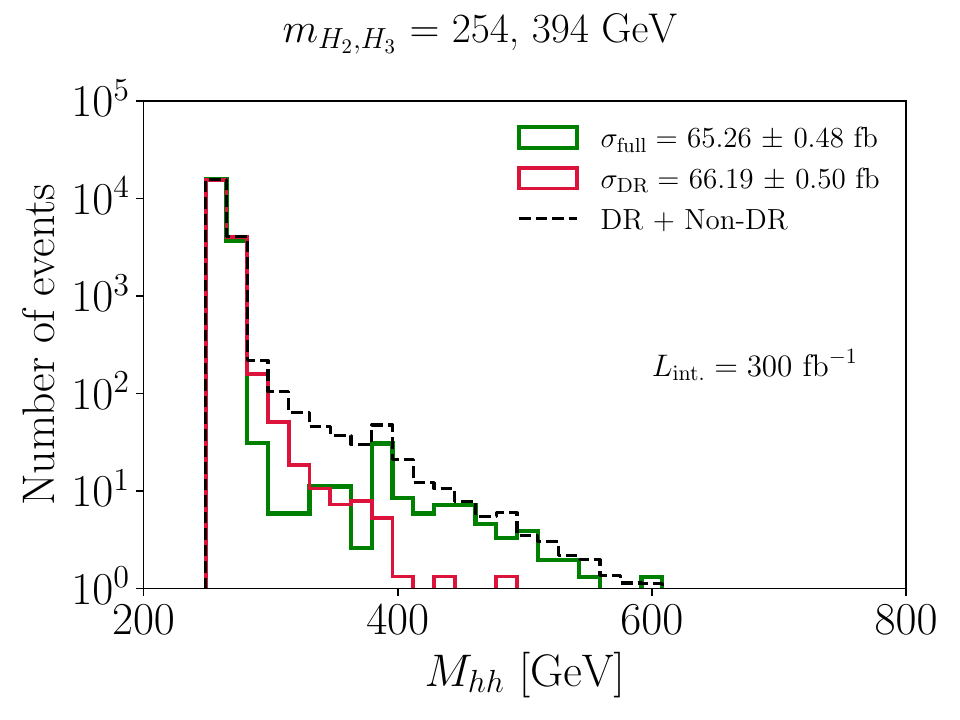}
\includegraphics[width=0.48\linewidth]{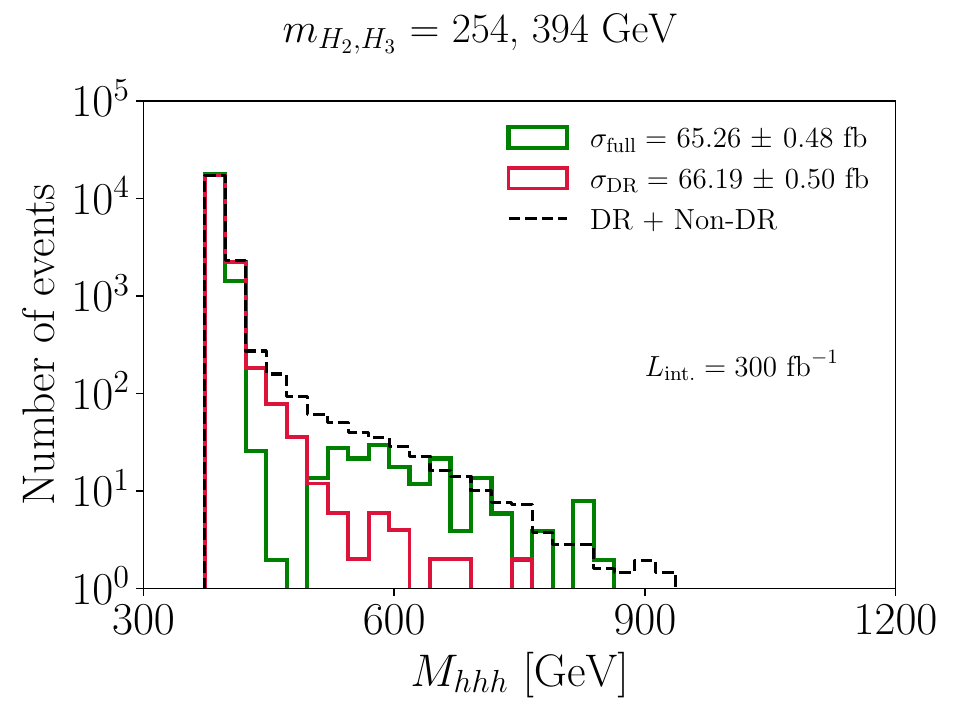}
\caption{The di- ($M_{hh}$) and tri-Higgs ($M_{hhh}$) invariant mass distributions for a mass configuration $m_{H_2,H_3} = 254,~394~\text{GeV}$ representing a scenario where most of the total cross-section (distributions shown in green) comes from the double-resonant process (distributions shown in red) and where the double-resonant cross-section is maximised (BP-1). Also illustrated on the plot is the combination of the double-resonant process and everything but the double-resonant process (shown as the black dashed line) elucidating the destructive interference effects.}
\label{fig:mhhh_254_394_100_max}
\end{figure}

We start by investigating a scenario where the mass of the second lightest scalar is about two times the mass of the 125 GeV Higgs boson and the mass of the heaviest scalar is about the sum of the masses of the second lightest and the 125 GeV Higgs boson, which is the region that yields the largest cross-sections of the double-resonance process, as well as its largest contributions to the total cross-section, as can be seen in Figs.~\ref{fig:xsec_masses},\ref{fig:xsecpercentages}.
In Fig.~\ref{fig:mhhh_254_394_100_max} one can see the distributions of the invariant masses of the di- and tri-Higgs boson system for our first benchmark point (BP-1), where the cross-section of the double-resonant process is maximised (with $\sigma_\text{DR} / \sigma_\text{full} 
\sim 1 $).
%%%%%%%%%%%%%%%%%%%%%%%%%%%%%%%%%%
%%%%%%%%%%%%%%%%%%%%%%%%%%%%%%%%%%

From the green lines, showing the distribution of the full process, and the red line, showing the distribution of only the double-resonant process, one can see that for this point the bulk of the full process can be well described by the double-resonant process. The di-Higgs boson invariant mass, shown on the left, peaks around $m_{H_2} \sim 254$~GeV, as one would expect, resulting from the two-body $H_2 \to h h $ decay. A smaller peak can be observed around $m_{H_3}\sim400$~GeV indicating the two-body decay of the heavy scalar $H_3 \to h h $ into two SM-like Higgs bosons.

The tri-Higgs boson invariant mass, on the right, peaks around $m_{H_3} \sim 400$~GeV, as also is expected from looking at the Feynman diagram of the double-resonant process in Fig.~\ref{fig:feyndiag1}. For the full process, the three-body decay from $H_3 \rightarrow hhh$ can also contribute here.
In both plots, one can observe negative interference by comparing the full process with the combination of the double-resonant process and everything but the double-resonant process, shown by the black dashed lines.
For all plots we show the sum of the double-resonant process and everything but the double-resonant process instead of showing the sum of the double-resonant process and the pure SM contribution, since the SM cross-section is about 100 - 1000 times lower ($\sim O(30-50$~ab)~\cite{deFlorian:2019app}) than the cross-section of the processes shown here.
In the left plot, for example, some negative interference can be seen around 300~GeV. This shows that in some cases, even though the full distributions can be well described by the double-resonant process, this does not mean that the double-resonant process alone dominates, but rather that negative interference pushes the full contribution down so that it is comparable to the double-resonant contribution.
In the right plot, negative interference around 450~GeV even leads to the contribution of the full process being lower than that the double-resonant contribution alone.

\begin{figure}[!t]
\centering
\includegraphics[width=0.48\linewidth]{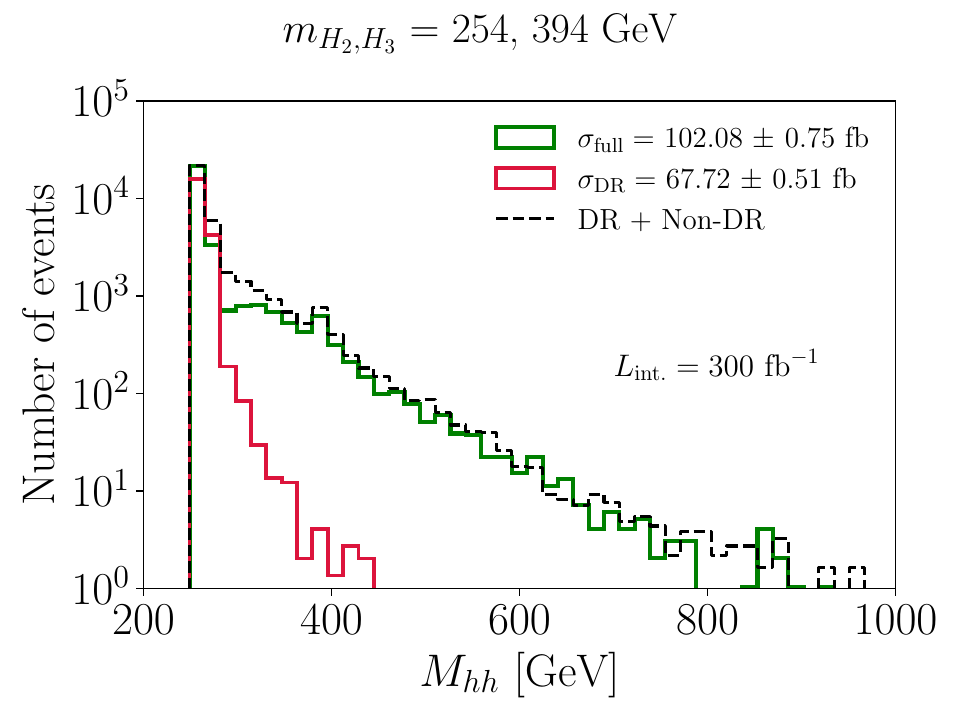}
\includegraphics[width=0.48\linewidth]{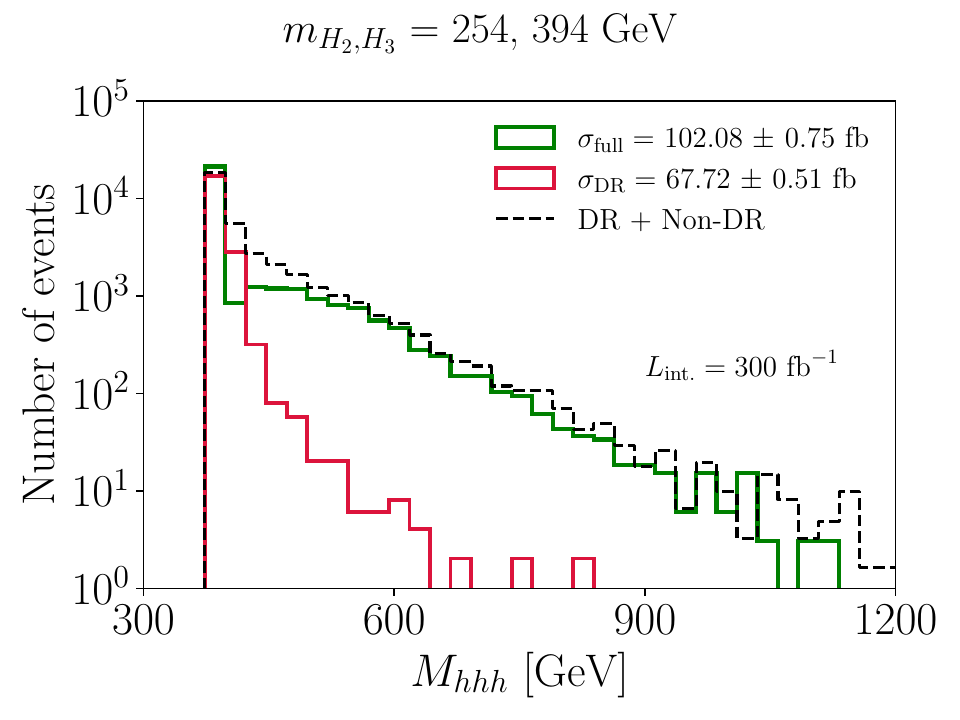}
\caption{The $M_{hh},~M_{hhh}$ distributions for a mass configuration $m_{H_2,H_3} = 254,~394~\text{GeV}$ representing a scenario {(BP-2)} where most of the total cross-section does not come fully from the double-resonant process even for a case where the double-resonant cross-section is maximised ($\sim 60$~fb). }
\label{fig:mhhh_254_394_100_min}
\end{figure}

In Fig.~\ref{fig:mhhh_254_394_100_min} the distributions of the invariant masses of the di- and tri-Higgs boson system are shown for another point (BP-2) with the same masses where the cross-section of the double-resonant process is maximised, however, for this point the contribution from the full process differs largely from the contribution of the double-resonant process only.

For this point the cross-section of the full process is much larger than for the previous point ($\sigma_\text{DR} / \sigma_\text{full} %\approx \frac{67.72}{102.08} 
\sim 0.66 $) indicating that even for a maximal cross-section for the double-resonance ($\sim 67~\text{fb}$), the full process can already have a significantly higher cross-section.
%%%%%%%%%%%%%%%%%%%%%%%%%%%%%%%%%%
%%%%%%%%%%%%%%%%%%%%%%%%%%%%%%%%%%

Again, the di- and tri-Higgs invariant masses peak at about $m_{H_2}\sim 250$~GeV and $m_{H_3} \sim 400$~GeV, respectively.
Some negative interference can be observed right after these peaks. The peaks, as in the first scenario, are well described by the double-resonant process. However, for the large tail above 300~GeV and above 500~GeV, for the di-and tri-Higgs invariant mass, respectively, the double-resonant contribution falls off very quickly and is not sufficient to describe the full process which falls off at a much lower rate due to the continuum contributions.
%%%%%%%%%%%%%%%%%%%%%%%%%%%%%%%%%%
\begin{figure}[!t]
\centering
\includegraphics[width=0.48\linewidth]{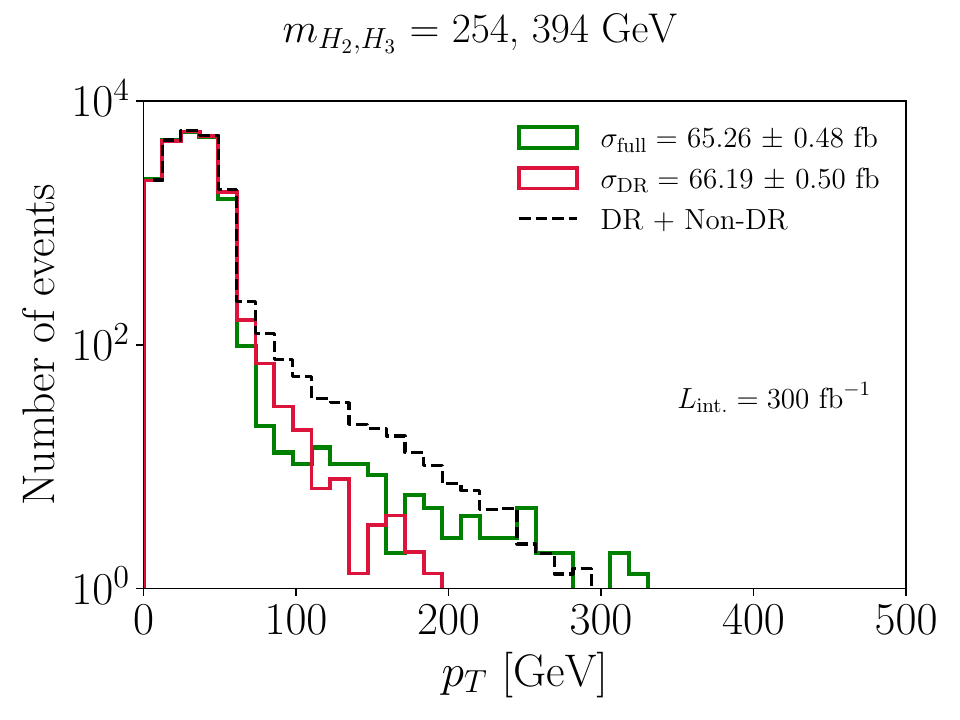}
\includegraphics[width=0.48\linewidth]{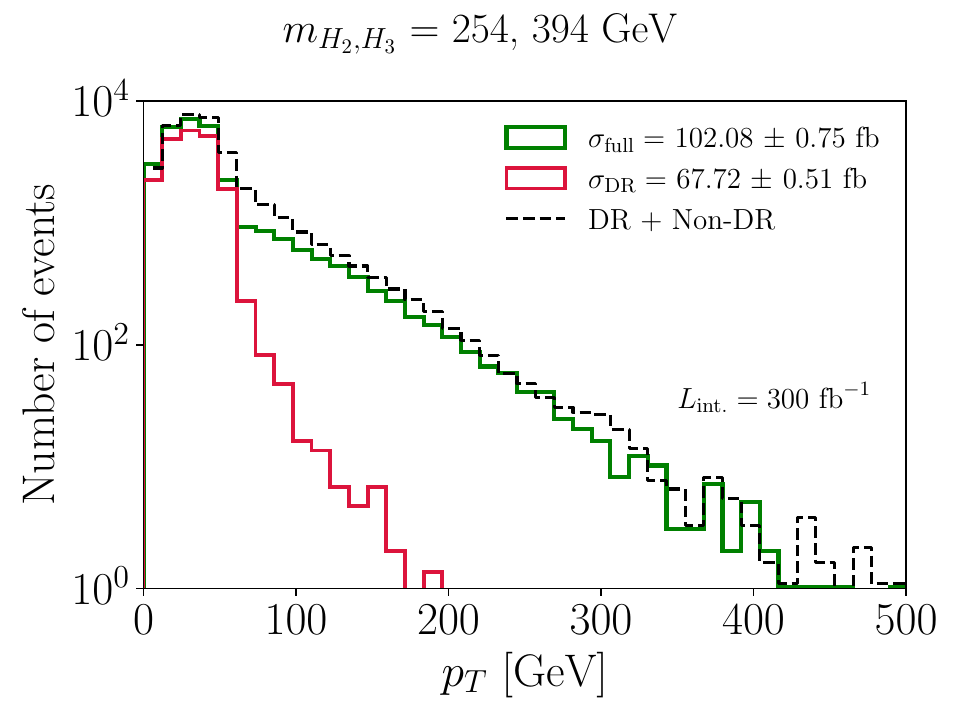}
\caption{The transverse momenta $(p_T)$ distributions for a mass configuration $m_{H_2,H_3} = 254,~394~\text{GeV}$ representing \textbf{(left)} a scenario (BP-1) where most of the total cross-section comes from the double-resonant process and \textbf{(right)} a scenario (BP-2) where most of the total cross-section doesn't come from the double-resonant process.}
\label{fig:pt_254_394_100}
\end{figure}
%%%%%%%%%%%%%%%%%%%%%%%%%%%%%%%%%%

Fig.~\ref{fig:pt_254_394_100} shows the $p_T$-distributions for the two points, discussed above. On the left, one can see the distribution for BP-1, where the full process and the double-resonant process lead to very similar contributions and on the right, one can see BP-2, where the double-resonant process alone is not sufficient to approximate the full process. For the left plot, as for the di- and tri-Higgs boson invariant mass, the full process is well-modelled by the double-resonant process up to values of about 200~GeV. Above that, statistical fluctuations can be seen. 
The transverse momentum peaks at low values and falls off rather quickly with some negative interference between about 100 and 200~GeV. The right plot, on the other hand, shows a longer tail in the full process for high values of the transverse momentum which is not modelled by the double-resonant process, which again falls off quickly for high momenta. One can again see the interference effects coming into the picture for both the points. This substantiates the fact that $p_T$ is also an important observable that one should consider when exploring kinematics in the triple Higgs frontier.

Since the scalar masses are identical for the two benchmark points considered above, one can expect that the observed differences in the kinematic distributions must originate from variations in the underlying coupling structure. However, attributing these features to a single parameter or interaction is, in general, not straightforward. The resulting kinematic behaviour arises from a highly non-trivial interplay among the various contributions arising in the full $g g \to hhh$ amplitude. A direct mapping between specific features in the distributions and individual parameters is not generically possible across the scanned parameter space, unless one resorts to a fully analytic decomposition of the amplitude, further underscoring the importance of studying the complete process through Monte-Carlo event generation instead of simplified descriptions.
%%%%%%%%%%%%%%%%%%%%%%%%%%%%%%%%%%
\begin{figure}[!t]
\centering
\includegraphics[width=0.48\linewidth]{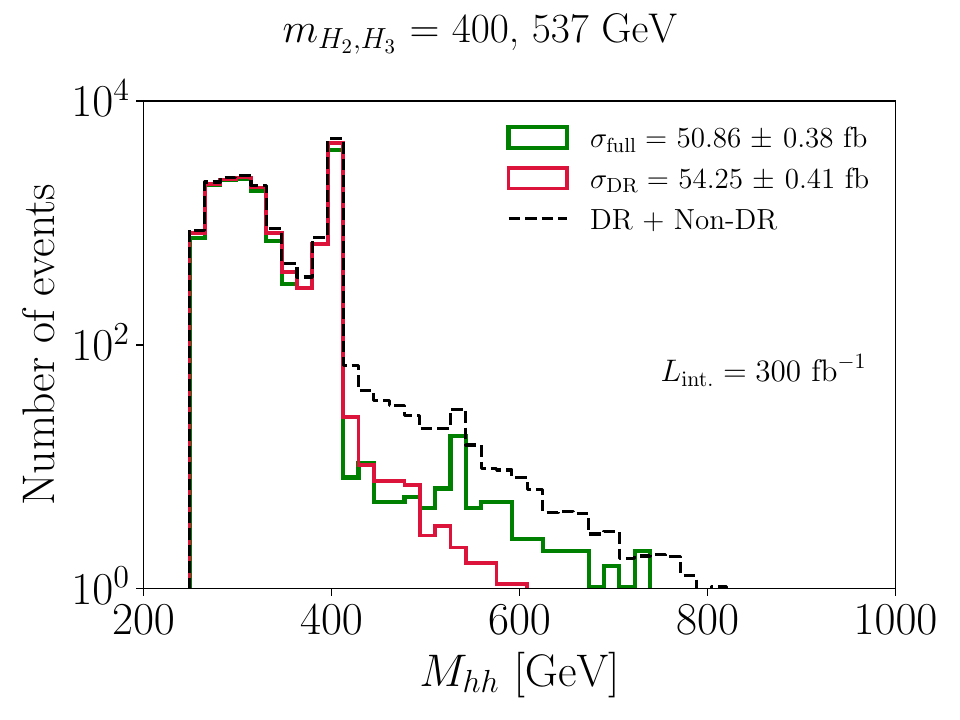}
\includegraphics[width=0.48\linewidth]{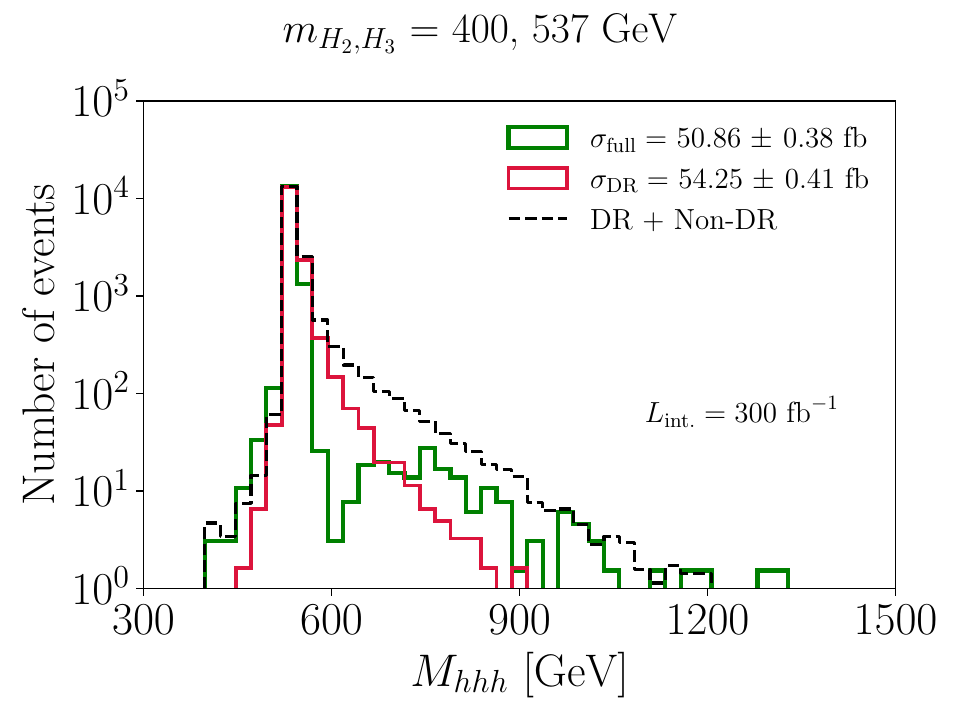}
\caption{The $M_{hh},~M_{hhh}$ distributions for a mass configuration $m_{H_2,H_3} = 400,~537~\text{GeV}$ representing a scenario (BP-3) where most of the total cross-section comes fully from the double-resonant process for a case where the double-resonant cross-section is maximised ($\sim 50$~fb). %{\bf tr what do you mean by even for a case} \wn{The `even' was a typo}
}
\label{fig:mhhh_400_537_100_maxed}
\end{figure}
%%%%%%%%%%%%%%%%%%%%%%%%%%%%%%%%%%

%%%%%%%%%%%%%%%%%%%%%%%%%%%%%%%%%%
\subsection{Heavier Mass Configurations}
\label{sec:heavymass}
%%%%%%%%%%%%%%%%%%%%%%%%%%%%%%%%%%
We now turn to exploring the kinematic distributions for a selection of additional mass benchmarks. We consider the mass configurations $m_{H_2,H_3} = 400,~537~\text{GeV}$ and $m_{H_2,H_3} = 600,~732~\text{GeV}$, since these exhibit a wider spread in the ratio of $\sigma_{\text{DR}}/\sigma_{\text{full}}$%\jz{full?}
, also including points where this ratio approaches unity (\textit{cf.} Fig.~\ref{fig:xsecpercentages}). For these benchmarks, the hierarchy $m_{H_2} > 3 m_h$ and $m_{H_3} \gtrsim m_{H_2} + m_h$ renders the additional 3-body decay $H_2 \to hhh$ kinematically accessible.
As a result, all relevant topologies can contribute to the full triple Higgs production process. One therefore expects these effects to leave visible imprints on the $M_{hh}$ and $M_{hhh}$ distributions, leading to a richer peak structure and making it more challenging to disentangle the relative contributions of the individual diagrams.

%%%%%%%%%%%%%%%%%%%%%%%%%%%%%%%%%%
\begin{figure}[!t]
\centering
\includegraphics[width=0.48\linewidth]{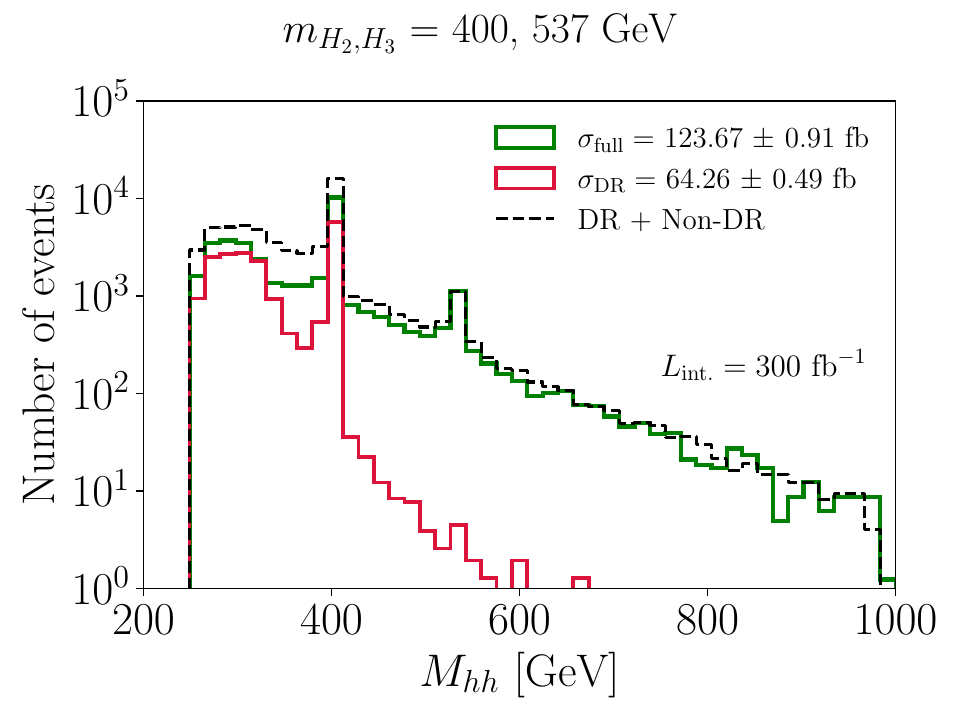}
\includegraphics[width=0.48\linewidth]{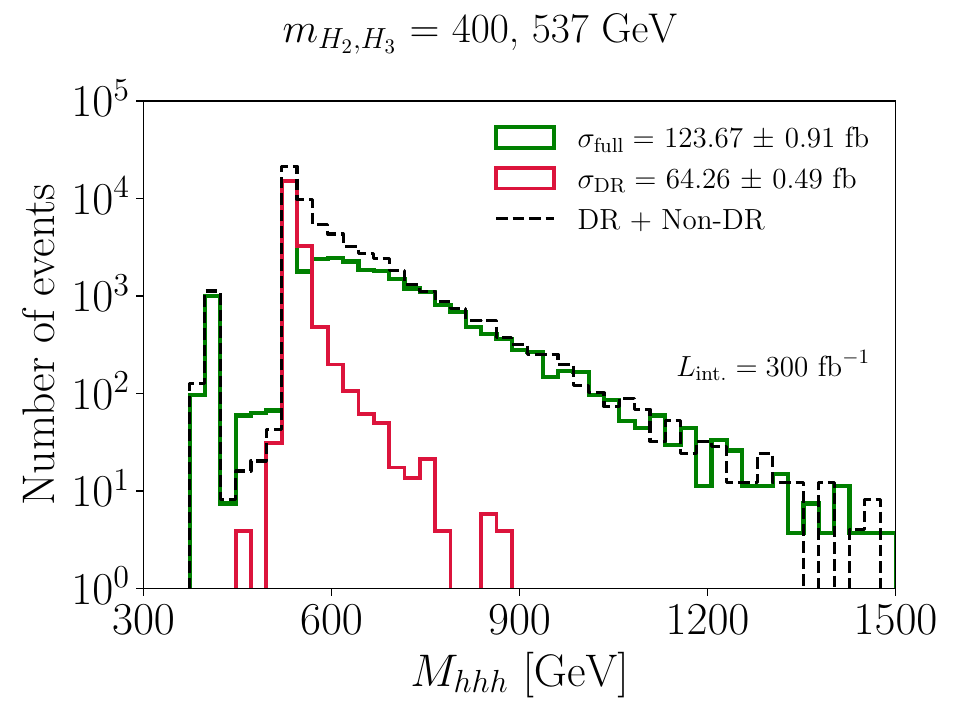}
\caption{The $M_{hh},~M_{hhh}$ distributions for a mass configuration $m_{H_2,H_3} = 400,~537~\text{GeV}$ representing a scenario (BP-4) where most of the total cross-section does not come fully from the double-resonant process even for a case where the double-resonant cross-section is maximised ($\sim 60$~fb). }
\label{fig:mhhh_400_537_100_min}
\end{figure}
%%%%%%%%%%%%%%%%%%%%%%%%%%%%%%%%%%
As in the previous mass configurations, we begin by examining a representative point (BP-3) with the kinematic configuration $m_{H_2,H_3} = 400,~537~\text{GeV}$, where the total cross-section is largely dominated by the double-resonant contribution where the latter is chosen at its maximal value ($\sim 55~\text{fb}$). The relevant kinematic distributions are shown in Fig.~\ref{fig:mhhh_400_537_100_maxed}. Similar to the lower-mass benchmarks, the double-resonant contribution provides, a good approximation to the full process for this point. The total cross-section is observed to be slightly smaller than the double-resonant contribution along, indicating the presence of destructive interference effects (this can be clearly observed again, comparing the dashed black line with the full process on Fig.~\ref{fig:mhhh_400_537_100_maxed}). The full process also exhibits slightly pronounced tails, arising from a combination of SM continuum contributions and subleading resonant topologies. Despite these effects, their overall impact remains limited, consistent with the observation here, that the total rate is largely captured by the double-resonant contribution.

This is however not always the case, as is illustrated by considering a benchmark point (BP-4) with identical masses, for which the double-resonant cross-section is again, maximal ($\sim 60$~fb), but the full process is not dominated by this particular topology. This is clearly illustrated in Fig.~\ref{fig:mhhh_400_537_100_min}, where the double-resonant contribution accounts for only about $\sim$50\% of the total cross-section. The kinematic distributions of $M_{hh},~M_{hhh}$ exhibit significant differences between the full process and the double-resonant contribution, with pronounced tails arising from the remaining topologies. Even in the presence of sizeable interference effects (comparing the dashed black and solid green distributions on Fig.~\ref{fig:mhhh_400_537_100_min}), the full cross-section can be substantially higher than the double-resonance contribution. 

%%%%%%%%%%%%%%%%%%%%%%%%%%%%%%%%%%
\begin{figure}[!t]
\centering
\includegraphics[width=0.48\linewidth]{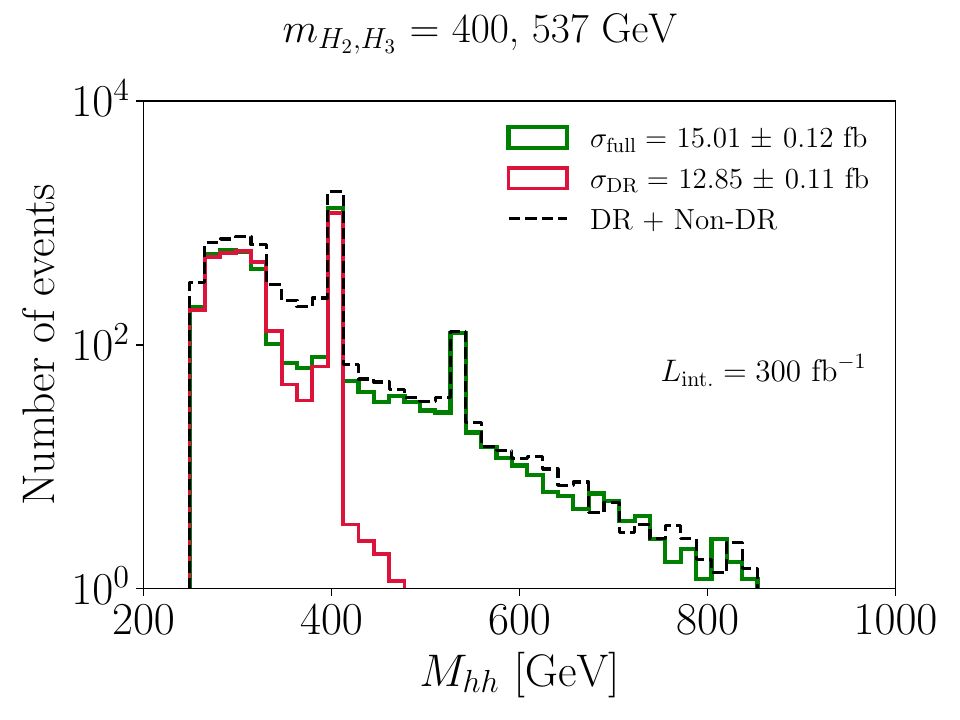}
\includegraphics[width=0.48\linewidth]{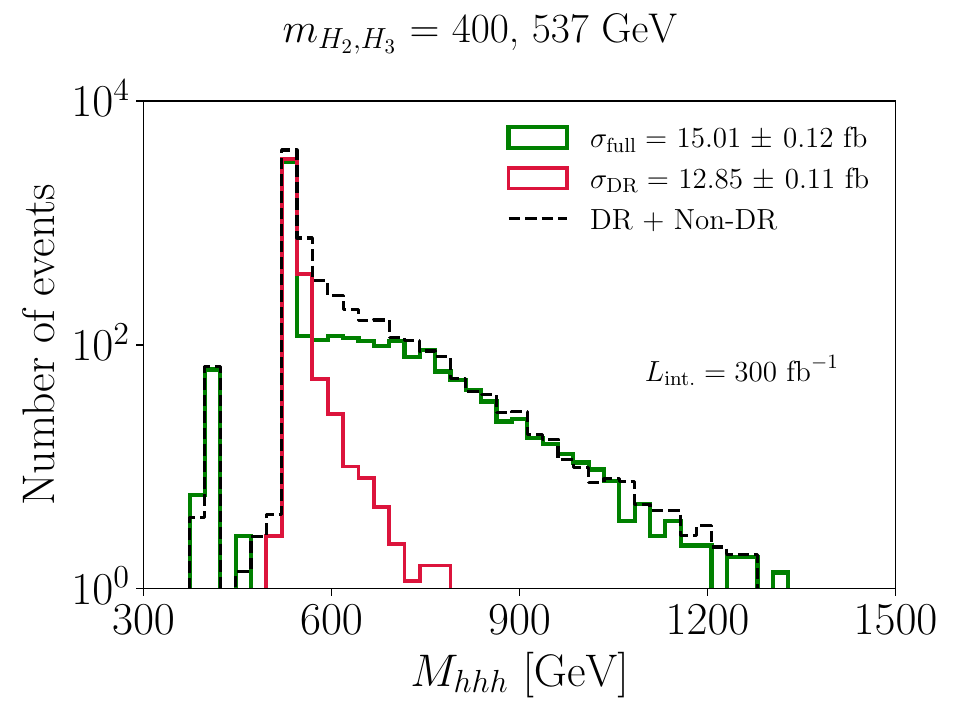}
\includegraphics[width=0.48\linewidth]{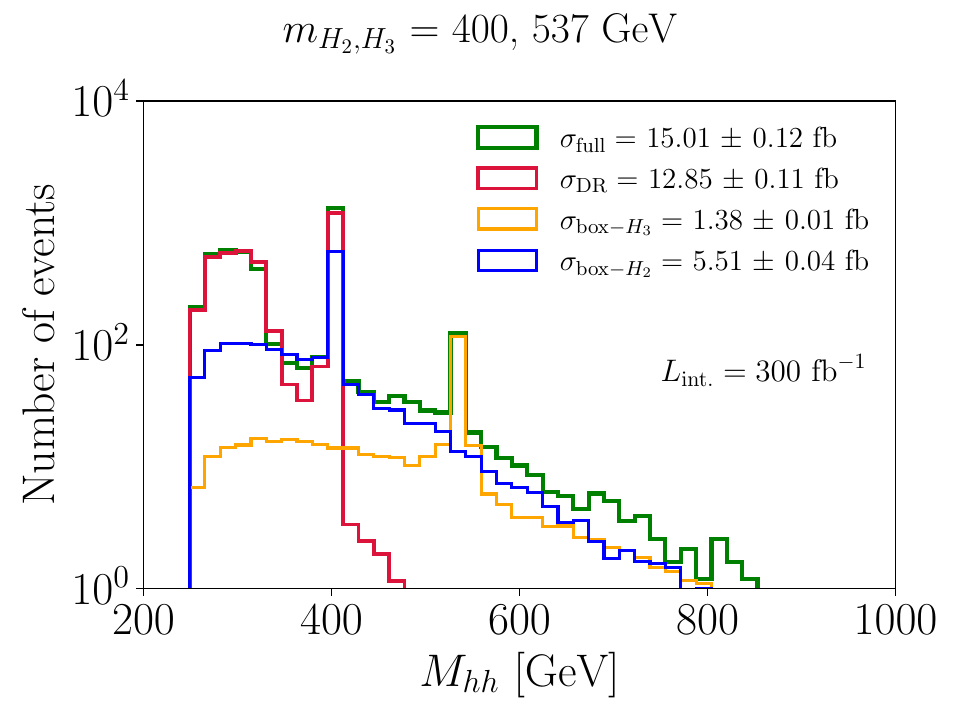}
\includegraphics[width=0.48\linewidth]{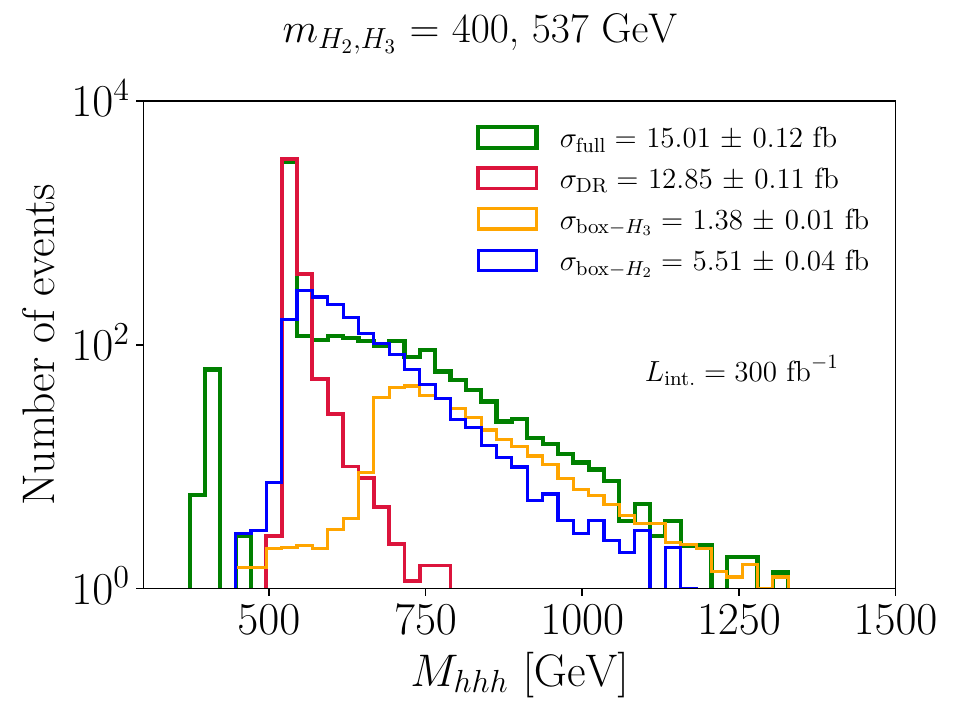}
\caption{The $M_{hh},~M_{hhh}$ distributions for a mass configuration $m_{H_2,H_3} = 400,~537~\text{GeV}$ representing a scenario (BP-5) where numerically, most of the total cross-section comes from the double-resonant process. The figures on the lower panel illustrate the various other Feynman topologies under consideration as described in the text.}
\label{fig:mhhh_400_537_100_max}
\end{figure}
%%%%%%%%%%%%%%%%%%%%%%%%%%%%%%%%%%

It is also not necessary that if the double-resonant cross-section, and the total cross-section is numerically similar, then the kinematic distributions for the full process would be well-modelled by the double-resonance alone. This behaviour is clearly illustrated for the another benchmark point (BP-5) with $m_{H_2, H_3} = 400,~537~\text{GeV}$, for which $\sigma_{\text{DR.}}/\sigma_{\text{full}}\sim 0.85$, indicating that the contribution from the double-resonance dominates numerically. The corresponding $M_{hh}$ and $M_{hhh}$ distributions are shown in Fig.~\ref{fig:mhhh_400_537_100_max}. The upper panel displays the results for the full process, the double-resonant topology, and the combination of the double-resonant contribution with the remainder, following the same conventions as before. Despite the numerical dominance of the double-resonance, the full process exhibits a pronounced tail in both distributions. The double-resonant topology, as expected, produces a clear peak at 400~GeV in the $M_{hh}$ distribution, corresponding to the two body decay of $H_2$, as well as a peak at 537~GeV associated with the subsequent decay of $H_3$. The full process, on the other hand, displays additional features, including a peak at 537~GeV in the $M_{hh}$ distribution arising from the two body decay of $H_3$, and a peak at 400~GeV in the $M_{hhh}$ distribution originating from the three body decay of $H_2$.

Further insight can be gained by examining selected individual contributions, as shown in the lower panel of Fig.~\ref{fig:mhhh_400_537_100_max}. The peak at 537~GeV in the $M_{hh}$ distribution can be attributed to the box diagram with $H_3$ propagating in one of the legs (shown in yellow), which subsequently decays into two Higgs bosons. Similarly, the box diagram involving $H_2$ (shown in blue) exhibits a distinct peak at 400~GeV in the $M_{hh}$ distribution, corresponding to the decay $H_2 \to h h$.

Finally, taken together, the plots in Fig.~\ref{fig:mhhh_400_537_100_max} indicate that destructive interference between the various contributions plays a significant role, leading to a total cross-section that is smaller than the naive sum of the individual components.
%%%%%%%%%%%%%%%%%%%%%%%%%%%%%%%%%%
\begin{figure}[!t]
\centering
\includegraphics[width=0.48\linewidth]{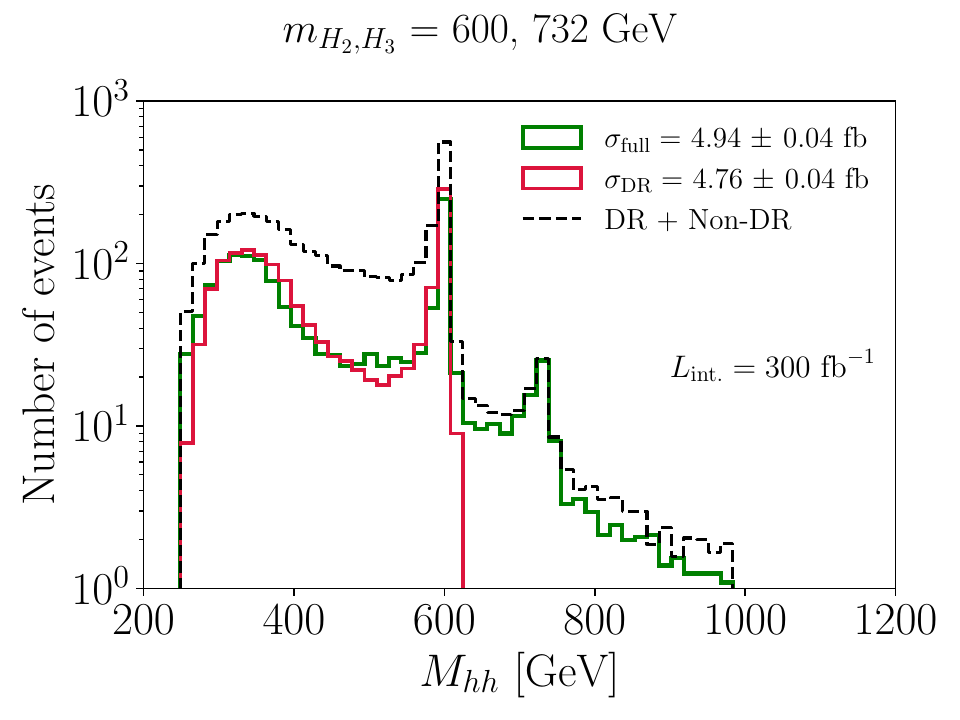}
\includegraphics[width=0.48\linewidth]{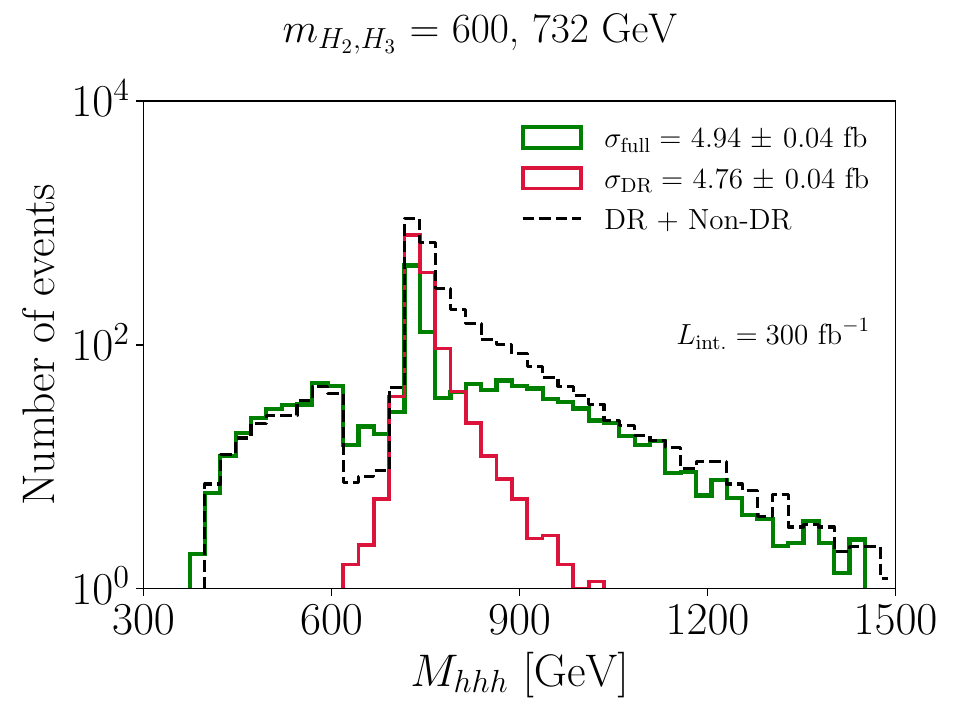}
\caption{The $M_{hh},~M_{hhh}$ distributions for a mass configuration $m_{H_2,H_3} = 600,~732~\text{GeV}$ representing a scenario (BP-6) where numerically, most of the total cross-section comes from the double-resonant process.}
\label{fig:mhhh_600_732_100_max}
\end{figure}
%%%%%%%%%%%%%%%%%%%%%%%%%%%%%%%%%%

%%%%%%%%%%%%%%%%%%%%%%%%%%%%%%%%%%
\begin{figure}[!t]
\centering
\includegraphics[width=0.48\linewidth]{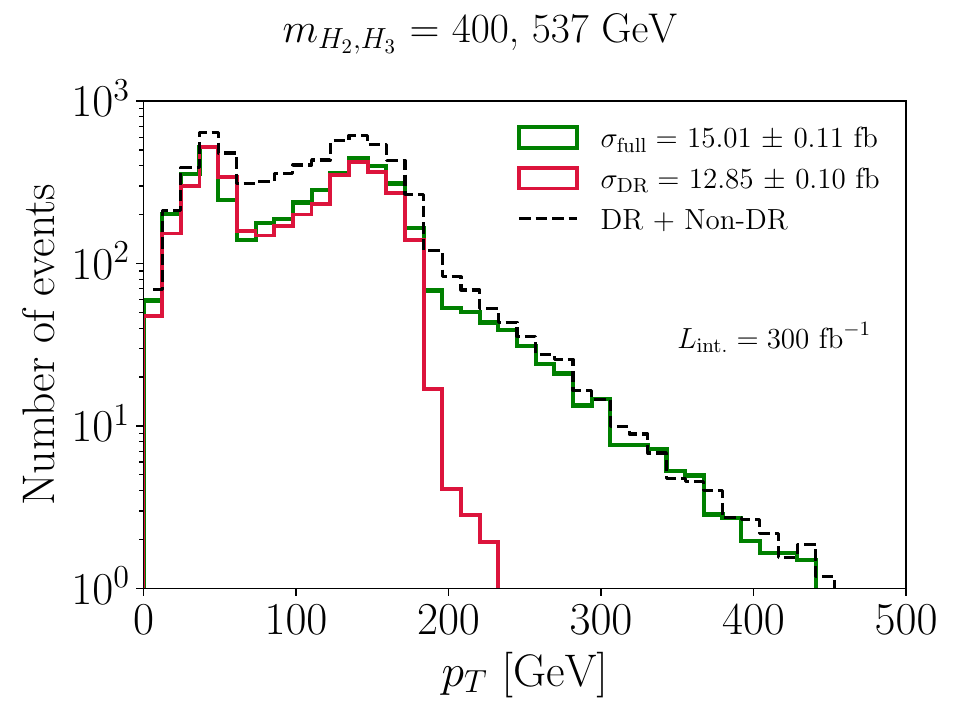}
\includegraphics[width=0.48\linewidth]{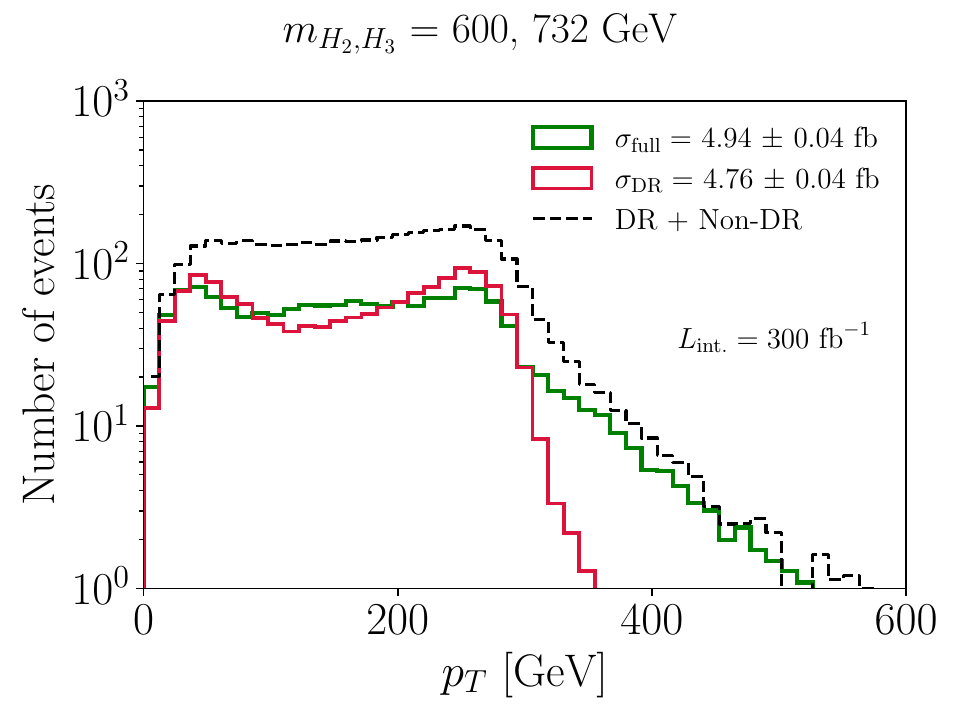}
\caption{The $p_T$ distributions for a mass configuration  \textbf{(left)} $m_{H_2,H_3} = 400,~537~\text{GeV}$ (BP-5) and \textbf{(right)} $m_{H_2,H_3} = 600,~732~\text{GeV}$ (BP-6), representing scenarios where numerically, most of the total cross-section comes from the double-resonant process.}
\label{fig:pt_400_and_600}
\end{figure}
%%%%%%%%%%%%%%%%%%%%%%%%%%%%%%%%%%

We next consider the benchmark characterised by $m_{H_2,H_3} = 600,~732~\text{GeV}$, choosing a point (BP-6) for which $\sigma_{\text{DR}}/\sigma_{\text{full}} \sim 1$. The corresponding distributions are shown in Fig.~\ref{fig:mhhh_600_732_100_max}. Despite the near unity ratio, a pronounced difference between the full process and the double-resonant contribution remains clearly visible. This further underlines that numerical agreement at the level of cross-sections does not faithfully reflect the underlying dynamical structure. The dashed curves in both panels highlight the sizeable destructive interference effects, which substantially reduce the full cross-section. This again emphasises that restricting the analysis to the double-resonant topology alone is insufficient to capture the full behaviour of the process.

Finally, one may consider the $p_T$ distributions for the benchmark points with heavier mass configurations (representative examples for BP-5 and BP-6 are shown Fig.~\ref{fig:pt_400_and_600}). While the location and spread of the peaks are largely dictated by the masses of the heavier resonances, significant differences between the double-resonant and full processes persist also in these observables, once again suggesting that $p_T$ is a useful additional handle in collider analyses for disentangling the underlying triple Higgs production mechanisms.

%%%%%%%%%%%%%%%%%%%%%%%%%%%%%%%%%%
\section{Conclusions}
\label{sec:conc}
%%%%%%%%%%%%%%%%%%%%%%%%%%%%%%%%%%
The exploration of multi-Higgs final states at the (HL-)LHC is a pivotal component of the BSM physics programme in the coming years. While di-Higgs production has already matured into a well-established probe, the study of triple Higgs final states has recently garnered increasing attention within both the theory and experimental communities. The inherent complexity of this process has motivated the development of theoretically well-grounded simplifications, as well as advances in analysis techniques. However, such simplifications can overlook important features arising from the rich kinematic structure of the full process. This provides the primary motivation for our study, in which we investigate the kinematic properties of resonant production of three Higgs bosons, in the presence of two additional neutral scalar degrees of freedom, in the CP-even N2HDM framework.

The central message from our results is that simplified descriptions based solely on factorised double-resonant topologies can be, in general, not sufficient to capture the underlying dynamics of $g g \to h h h$. Even though they can provide useful intuition, they can lead to misleading conclusions when used as a proxy to the full process. This limitation further persists across all kinematic regimes in which the double-resonant configuration is kinematically accessible, including both near-threshold regions and scenarios with widely separated mass scales. Even in cases where resonant enhancements are present, offshell contributions and interference effects can play a significant role. Destructive interference, which we observe for all considered benchmark points, in particular, can substantially modify the total cross-section, implying numerical similarity between double-resonant and full rates does not necessarily indicate dominance of the former.

Although our analysis is performed within the N2HDM scalar spectra, the underlying features are expected to be generic to renormalisable scalar extensions of the SM with three (or more) additional degrees of freedom, such as C2HDM~\cite{Fontes:2017zfn}, or triplet scenarios like the Georgi-Machacek model~\cite{Georgi:1985nv}. Such scenarios result in correlations among masses and couplings, together with the presence of multiple contributing topologies, naturally leading to deviations from simplified descriptions based on sequential onshell decays.

From a phenomenological perspective, invariant mass observables $M_{hh},~M_{hhh}$ provide the most direct sensitivity to the underlying production mechanisms and can partially disentangle the individual contributing topologies. The $p_T$-distributions also exhibit discernible differences between the double-resonant approximation and the full process. These observations have direct implications for analysis strategies, in particular, the dominant role of the invariant mass variables suggests that they should be included explicitly, alongside four-momentum information, in multivariate approaches such as boosted decision trees, and graph neural networks. 

Our results therefore highlight the importance of going beyond simplified descriptions in the study of triple Higgs production. A faithful interpretation of potential signals requires accounting for the full kinematic structure of the process, including interference and offshell effects. As the experimental programme continues to advance toward higher luminosities and increasingly sophisticated analysis techniques, such comprehensive approaches will be essential for fully exploiting the potential of multi-Higgs final states as probes of the scalar sector and BSM physics.

%%%%%%%%%%%%%%%%%%%%%%%%%%%%%%%%%%
%\acknowledgments
\section*{Acknowledgements}
W.N. thanks Christoph Englert for the \texttt{FeynRules} implementation of the N2HDM scalar extension. 
We acknowledge Rui Santos for helpful discussions.
TR acknowledges financial support from the Croatian Science Foundation (HRZZ)
project ``Beyond the Standard Model discovery and Standard Model precision at
LHC Run III", IP-2022-10-2520, and thanks the University of Hamburg as well as DESY Theory group for their hospitality while this work was initiated.
All authors acknowledge support by the Deutsche Forschungsgemeinschaft (DFG, German Research Foundation) under Germany's Excellence Strategy --- EXC 2121 ``Quantum Universe'' --- 390833306. This work has been partially funded by the Deutsche Forschungsgemeinschaft (DFG, German Research Foundation) --- 491245950. 

%%%%%%%%%%%%%%%%%%%
\appendix
%%%%%%%%%%%%%%%%%%%%%%%%%%%%%%%%%%%%
\section{Angular Variables}
\label{app:angular}
%%%%%%%%%%%%%%%%%%%%%%%%%%%%%%%%%%%%
In this appendix, we comment on the relevance of angular observables, in particular, the pseudorapidity $\eta$ and azimuthal angle $\phi$ of the final-state Higgs bosons, for distinguishing between the full process and the double-resonant contribution. As a representative example, we consider the benchmark point {BP-5} with masses $m_{H_2,~H_3} = 400,~537~\text{GeV}$ and $\sigma_{\text{DR}}/\sigma_{\text{full}} \sim 0.85$. The corresponding $M_{hh}-,~M_{hhh}-$distributions for this point (shown in Fig.~\ref{fig:mhhh_400_537_100_max}) have already been discussed in detail in Sec.~\ref{sec:heavymass}, where it was demonstrated that even when the double-resonant cross-section is numerically close to the total cross-section, the kinematic distributions can exhibit significant differences. 

\begin{figure}[!t]
\centering
\includegraphics[width=0.48\linewidth]{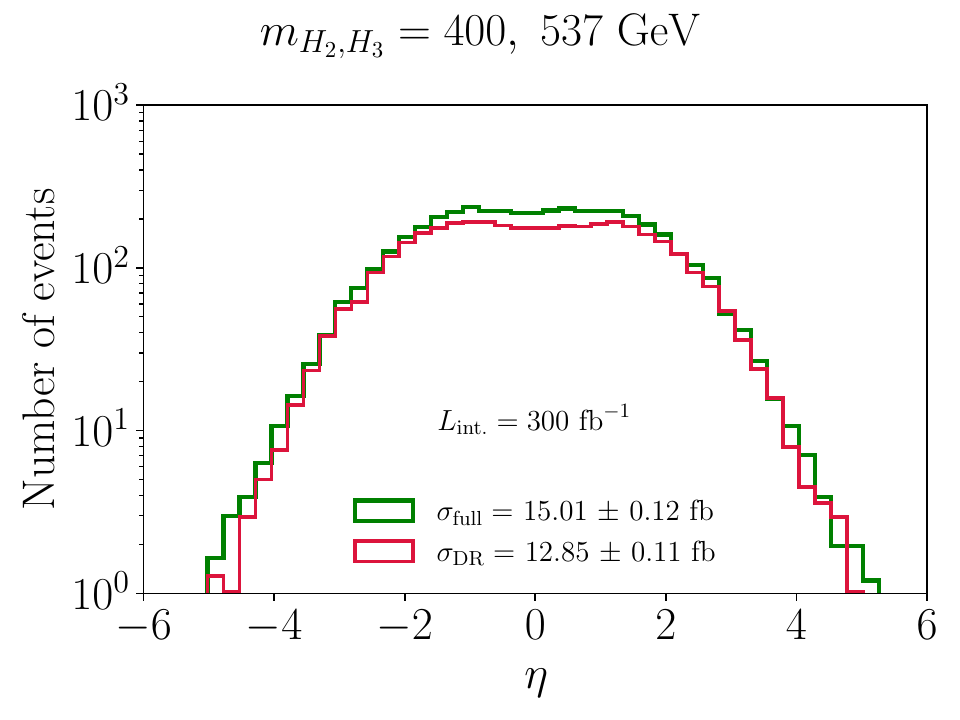}
\includegraphics[width=0.48\linewidth]{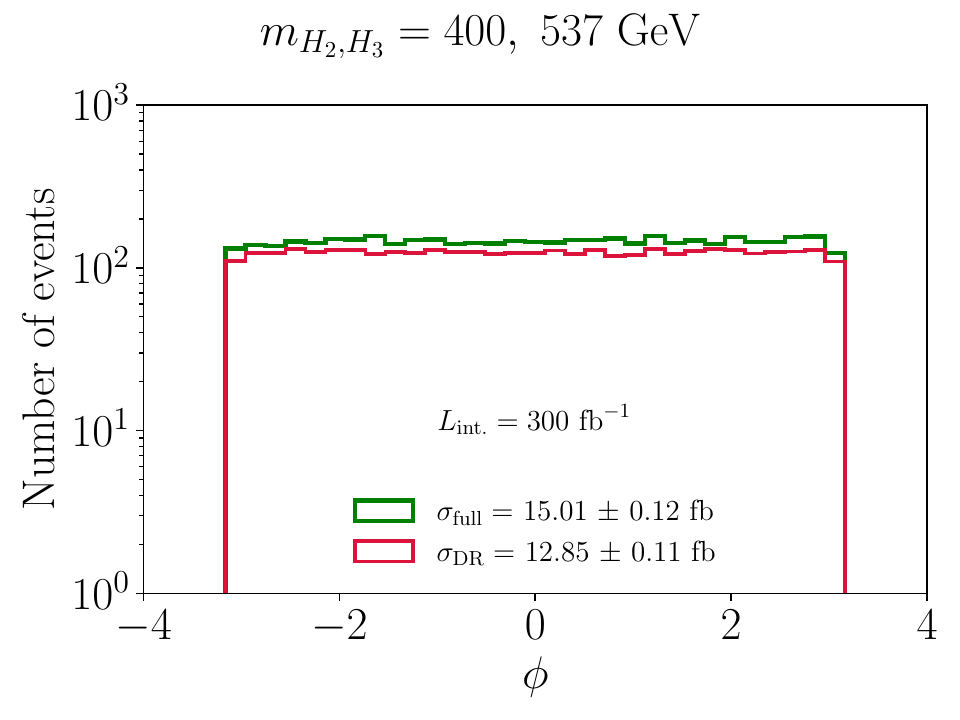}
\caption{Distributions of the angular observables $\eta$ and $\phi$ for the full, and the double-resonant processes. }
\label{fig:etaphi}
\end{figure}

The angular variables for this benchmark are shown in Fig.~\ref{fig:etaphi}, where the green distribution represents the full process, and the red distribution corresponds to the double-resonance contribution. In contrast to the invariant mass and transverse momentum observables, the shapes of the $\eta-$ and $\phi-$ distributions are very similar for both considered cases, with no discernible structural differences.
%%%%%%%%%%%%%%%%%%%%%%%%%%%%%%%%%%%%
%%%%%%%%%%%%%%%%%%%%%%%%%%%%%%%%%%%%

We have verified that this behaviour persists across other benchmark points with different mass spectra and coupling configurations. While normalising these distributions to the cross-sections leads to different event yields per bin, this simply reflects the overall rate differences between the full process and the double-resonant contribution. At the level of our parton-level analysis, we therefore conclude that the kinematic features identified above have limited impact on angular observables. 

%%%%%%%%%%%%%%%%%%%
\bibliographystyle{JHEP}
\bibliography{references}
%%%%%%%%%%%%%%%%%%%

\end{document}